\documentclass[prd,showpacs,superscriptaddress,floatfix,nofootinbib,10pt]{revtex4-2}
\usepackage{times}
\usepackage{amssymb,amsbsy,amsmath,amsfonts}
\usepackage{graphicx}
\usepackage{float}
\usepackage{color}
\usepackage{orcidlink}
\usepackage{morefloats}
\usepackage{rotating}
\usepackage{srcltx}
\usepackage{slashed}
\usepackage{multirow}
\usepackage{verbatim}
\usepackage{hyperref}
\usepackage{tabularx}
\usepackage{bm}
\allowdisplaybreaks[4]

\usepackage{bbding}
\usepackage{threeparttable}

\usepackage{mathrsfs}

\newcommand{\PreserveBackslash}[1]{\let\temp=\\#1\let\\=\temp}
\newcolumntype{C}[1]{>{\PreserveBackslash\centering}p{#1}}
\newcolumntype{R}[1]{>{\PreserveBackslash\raggedleft}p{#1}}
\newcolumntype{L}[1]{>{\PreserveBackslash\raggedright}p{#1}}
\graphicspath{{figures/}}
\allowdisplaybreaks[0]
\newcommand{\At}{\widetilde A}
\newcommand{\Bt}{\widetilde B}
\newcommand{\Ct}{\widetilde C}
\newcommand{\dd}{\mathrm d}
\newcommand{\MM}{\mathrm{MM}}
\newcommand{\MB}{\mathrm{MB}}
\newcommand{\MbarB}{\mathrm{M\bar B}}
\newcommand{\mix}{\mathrm{mix}}
\newcommand{\Tree}{\mathrm{Tree}}
\newcommand{\pip}{\pi^+\pi^-}
\newcommand{\pin}{\pi^0\pi^0}
\newcommand{\pieta}{\pi^0\eta}
\newcommand{\kkc}{K^+K^-}
\newcommand{\kkn}{K^0\bar K^0}
\newcommand{\pL}{\pi^0\Lambda}
\newcommand{\eL}{\eta\Lambda}
\newcommand{\pS}{\pi^0\Sigma^0}
\newcommand{\eS}{\eta\Sigma^0}
\newcommand{\psm}{\pi^+\Sigma^-}
\newcommand{\msp}{\pi^-\Sigma^+}
\newcommand{\kp}{K^-p}
\newcommand{\kn}{\bar K^0n}
\newcommand{\kx}{K^+\Xi^-}
\newcommand{\kz}{K^0\Xi^0}
\newcommand{\GT}[2]{G_{#1}T_{#1,#2}}
\newcommand{\MBGT}[2]{G_{#1}T_{#1,#2}}

\begin{document}

\title{\texorpdfstring{Isospin violation and final-state interactions in\\
$J/\psi,\ \psi(3686)\to\Lambda\bar\Lambda\pi^0\eta$}{Isospin violation and final-state interactions in charmonium four-body decays}}

\author{Jing Song\orcidlink{0000-0003-3789-7504}}
\email[]{Song-Jing@buaa.edu.cn}
\affiliation{Center for Theoretical Physics, School of Physics and Optoelectronic Engineering, Hainan University, Haikou 570228, China}
\affiliation{School of Physics, Beihang University, Beijing, 102206, China}
\affiliation{Departamento de Física Teórica and IFIC, Centro Mixto Universidad de Valencia-CSIC Institutos de Investigación de Paterna, 46071 Valencia, Spain}

\author{Yu-Shan Ren\orcidlink{https://orcid.org/0009-0001-4911-3829}}
\email[]{Yushan.Ren@ific.uv.es}
\affiliation{School of Physics, University of Electronic Science and Technology of China, \\Chengdu 610054, China}
\affiliation{Departamento de Física Teórica and IFIC, Centro Mixto Universidad de Valencia-CSIC Institutos de Investigación de Paterna, 46071 Valencia, Spain}
\vspace{0.5cm}

\author{Eulogio Oset\orcidlink{0000-0002-4462-7919}}%
\email[]{oset@ific.uv.es}
\affiliation{Departamento de Física Teórica and IFIC, Centro Mixto Universidad de Valencia-CSIC Institutos de Investigación de Paterna, 46071 Valencia, Spain}	
\affiliation{Department of Physics, Guangxi Normal University, Guilin 541004, China}

\begin{abstract}
We study the \(J/\psi\) and \(\psi(1686)\) decays to \(\bar{\Lambda}\pi^{0}\eta\Lambda\), exploiting the singlet \(\mathrm{SU}(3)\) structure in the light quarks of the \(J/\psi\) and \(\psi(1686)\) states, constructing all the singlet structures involving \(\bar{B}PPB\), where \(B\) are baryons and \(P\) pseudoscalar mesons of the \(\mathrm{SU}(3)\) octets. The reaction violates isospin, and is only made possible through final state interaction of meson baryon and \(\pi^{0}\)-\(\eta\) mixing. In the \(\psi(1686)\) decay, that has more phase space available, we find signals of the dynamically generated resonances, \(a_{0}(980)\) in the \(\pi^{0}\eta\) mass distribution and of the \(\Sigma(1430)\) and \(\Lambda(1670)\) in the \(\pi^{0}\Lambda\) mass distribution. The peak of the \(\Sigma(1430)\) shows up with large strength in the mass distribution, offering a valuable complement to the Belle reaction on \(\Lambda_{c}\to\Lambda\pi^{+}\pi^{+}\pi^{-}\) reaction where the \(\Sigma(1430)\) was observed as a very small peak over the background.
\end{abstract}

\maketitle

\section{Introduction}
\label{sec:introduction}

Hadronic decays of the $J/\psi$ and $\psi(3686)$ provide a useful environment to investigate low-energy hadron interactions and the formation of hadronic resonances~\cite{Oset:2016lyh,Guo:2017jvc}. Three-body hadronic decays have been extensively studied to obtain information on the interactions of final hadron pairs~\cite{Oset:2016lyh,Guo:2017jvc}. The extension from three-body to four-body decays introduces additional complexity because several pairs of particles can interact and their contributions can interfere with each other. At the same time, four-body decays allow one to investigate several two-body subsystems through different invariant-mass distributions in the same reaction. Experimentally, four-body decay channels such as $J/\psi\to\Lambda\bar\Lambda\pi^+\pi^-$, $D_s^+\to\pi^+\pi^+\pi^-\eta$, and $D_s^+\to K_S^0K^-\pi^+\pi^+$ have been studied~\cite{BESII:2012oco,BESIII:2021aza,BESIII:2021dot}. More recently, the amplitude analyses of $D_s^+\to\pi^+\pi^+\pi^-\pi^0$ and $D_s^+\to\pi^+\pi^0\pi^0\eta$ have provided further information on the contributions of intermediate states in four-body decays~\cite{BESIII:2024muy,BESIII:2026mtz}. The role of pairwise rescattering in four-body mass distributions has been illustrated, for example, in the study of $D_s^+\to\pi^+\pi^+\pi^-\eta$~\cite{Song:2022kac}. A related coupled-channel study of $D_s^+\to\pi^+\pi^0\pi^0\eta$ has also been performed in~\cite{He:2026iiy} and of $D^{+} \to \bar{K}^{0} \pi^{+} \pi^{0} \pi^{0}$ in~\cite{Yang:2026bhm}. Other examples are $X(3872)\to J/\psi \pi^+\pi^-\pi^0$ in~\cite{Wang:2022vjm} and $\Lambda_c\to \pi^+\pi^0\pi^-\Sigma^+$ in~\cite{Xie:2018gbi}.

Since the $J/\psi$ and $\psi(3686)$ are singlets under light-flavor SU(3), the dominant strong-interaction production mechanism conserves isospin. The $\Lambda\bar\Lambda$ pair has isospin zero, whereas the $\pi^0\eta$ pair has isospin one. Consequently, the $\Lambda\bar\Lambda\pi^0\eta$ final state offers a particularly interesting opportunity to study isospin-breaking effects. The BESIII Collaboration has measured the isospin-violating $J/\psi\to\Lambda\bar\Lambda\pi^0$ decay, together with the isospin-conserving $J/\psi\to\Lambda\bar\Lambda\eta$ and $\psi(3686)\to\Lambda\bar\Lambda\eta$ reactions~\cite{BESIII:2012jve}. A later analysis also reported evidence for $\psi(3686)\to\Lambda\bar\Lambda\pi^0$~\cite{BESIII:2022cxi}. These studies illustrate how processes suppressed by isospin violation can be sensitive to the dynamics of the final-state hadrons. In the present four-body reaction, the $\pi^0\eta$, $\pi^0\Lambda$ ($\pi^0\bar\Lambda$), and $\eta\Lambda$ ($\eta\bar\Lambda$) subsystems are particularly relevant.

The $\pi^0\eta$ system is of special interest because the $a_0(980)$ is strongly affected by the coupled $\pi\eta$ and $K\bar K$ interactions. Within the chiral unitary approach, the light scalar mesons emerge from the nonperturbative interaction of pairs of pseudoscalar mesons~\cite{Oller:1997ti,Hanhart:2007bd,Guo:2017jvc}. The proximity of the $K^+K^-$ and $K^0\bar K^0$ thresholds to the $a_0(980)$ region makes the line shape particularly sensitive to coupled-channel dynamics~\cite{Hanhart:2007bd,Lin:2021isc}. A decay producing a $\pi^0\eta$ pair in an isospin forbidden decay mode can therefore provide information complementary to that obtained from  other hadronic decays conserving isospin.

Another relevant subsystem is $\pi\Lambda$, which has isospin $I=1$ and couples to $\bar K N$, $\pi\Sigma$, $\eta\Sigma$, and $K\Xi$ channels. The coupled-channel treatment of the strangeness $S=-1$ meson--baryon interaction has a long history in the chiral unitary approach~\cite{Oset:1997it,Oller:2000fj,Jido:2003cb,Hyodo:2011ur,Garcia-Recio:2003ejq,Feijoo:2018den}.
The Belle Collaboration observed pronounced structures in the $\Lambda\pi^+$ and $\Lambda\pi^-$ mass distributions near the $I=1$ $\bar K N$ thresholds in $\Lambda_c^+\to\Lambda\pi^+\pi^+\pi^-$ decay~\cite{Belle:2022ywa}. The structure, often referred to as the $\Sigma(1430)$, has subsequently been investigated through coupled-channel correlation functions and decay mechanisms~\cite{Li:2024tvo,Li:2025yad}. A related analysis of $J/\psi\to\Lambda\bar\Lambda\pi^0$ also found an enhancement in this region induced by isospin-breaking meson--baryon rescattering~\cite{Ren:2026ila}. Triangle-singularity mechanisms have also been explored as a possible source of resonancelike structures in the isospin-violating $J/\psi\to\Lambda\bar\Lambda\pi$ reaction~\cite{Huang:2024oai}. These developments make the $\pi^0\Lambda$ and $\pi^0\bar\Lambda$ mass distributions attractive observables for examining the $I=1$ dynamics without presupposing that every threshold enhancement represents an isolated resonance pole.

The $\eta\Lambda$ interaction offers a complementary probe of the $I=0$ strangeness $S=-1$ sector. The $\Lambda(1670)$ has been described as a dynamically generated state in chiral coupled-channel studies, with an important role played by the $\eta\Lambda$ and other meson--baryon channels~\cite{Oset:2001cn,Jido:2003cb,Xie:2016evi,Wang:2022nac,Duan:2024czu,Zhang:2026oml}. 
The proximity of the $\eta\Lambda$ threshold to this resonance makes its line shape sensitive to the corresponding scattering amplitude. Recent analyses of experimental information have constrained the $\eta\Lambda$ scattering length, effective range, and the pole associated with the $\Lambda(1670)$~\cite{Duan:2024okk}. The BESIII Collaboration found that the $\eta\Lambda$ threshold enhancement in $\psi(3686)\to\Lambda\bar\Lambda\eta$ is compatible with a contribution from the $\Lambda(1670)$ resonance~\cite{BESIII:2022cxi}. In a related theoretical study of this decay, the measured $\eta\Lambda$ and $\eta\bar\Lambda$ mass distributions were reproduced using coupled-channel final-state interactions that generate the $\Lambda(1670)$~\cite{Ikeno:2026zpv}. The $\Lambda(1670)$ has also been identified in a partial-wave analysis of $\psi(3686)\to\Lambda\bar\Sigma^0\pi^0+\mathrm{c.c.}$~\cite{BESIII:2024jgy}. More recently, isospin violation in $J/\psi\to\Lambda\bar\Sigma^0\eta$ has been investigated using SU(3) production sources followed by meson--baryon final-state interactions~\cite{Dai:2026zqn}; related amplitude analyses of $J/\psi\to\Lambda\bar\Sigma^0\eta+\mathrm{c.c.}$ have identified contributions from the $\Lambda(1670)$ in the $\eta\Lambda$ subsystem~\cite{BESIII:2026ovz}. Thus, the same four-body final state can probe the $a_0(980)$ region in $\pi^0\eta$, the $I=1$ $\pi\Lambda$ interaction near the $\Sigma(1430)$ region, and the $I=0$ $\eta\Lambda$ interaction associated with the $\Lambda(1670)$. The corresponding meson--antibaryon interactions provide charge-conjugated contributions to the mass distributions.

However, the existence of these coupled-channel interactions does not change the isospin suppression of the primary $J/\psi,\psi(3686)\to\Lambda\bar\Lambda\pi^0\eta$ decay. There are different sources of isospin violation in low-energy QCD, notably the difference between the up- and down-quark masses and electromagnetic interactions~\cite{Gasser:1983yg,Gasser:1984gg}.
At the hadronic level, isospin breaking arises from the mass differences between charged and neutral hadrons and from $\pi^0$--$\eta$ mixing.The different charged-channel thresholds can prevent contributions from canceling exactly even when the primary production source conserves isospin. This mechanism can become particularly important near a two-body threshold, where small mass differences may lead to appreciable changes in the scattering amplitudes. A familiar example is the $a_0(980)$--$f_0(980)$ mixing driven by the difference between the charged- and neutral-kaon thresholds~\cite{Hanhart:2007bd}, which has been observed experimentally by the BESIII Collaboration~\cite{BESIII:2018ozj}. The interplay between isospin breaking in the production loops and in the coupled-channel meson--meson amplitudes has also been studied in $D_s^+\to a_0(980)e^+\nu_e$ decay~\cite{Lin:2021isc}. In the reactions considered here, both the physical mass splittings in the intermediate channels and the $\pi^0$--$\eta$ mixing contributions have to be taken into account. 

For the $\Lambda\bar\Lambda\pi^0\eta$ final state, we consider meson--meson, meson--baryon, and meson--antibaryon rescattering, together with the effects of $\pi^0$--$\eta$ mixing. A description based on coupled-channel scattering amplitudes provides a common dynamical framework for these contributions, while the different production and rescattering mechanisms are added coherently. Such a treatment is important because an enhancement in one invariant-mass distribution can also be reflected in another distribution after integration over the four-body phase space.

In the present work, we study the isospin-violating four-body decays $J/\psi,\psi(3686)\to\Lambda\bar\Lambda\pi^0\eta$. We start from an SU(3)-invariant production source and account for the final-state interactions of the relevant meson--meson, meson--baryon, and meson--antibaryon pairs using the chiral unitary approach. Isospin breaking is introduced through the physical charged-channel masses and $\pi^0$--$\eta$ mixing, and all the resulting contributions are summed coherently. We then calculate the six two-particle invariant-mass distributions for each charmonium state and compare their shapes with the corresponding pure phase-space distributions. An important distinction between the two decays arises from their different kinematic ranges: the $\pi^0\eta$ invariant mass in $J/\psi$ decay cannot reach the $a_0(980)$ region, whereas $\psi(3686)$ decay covers this region as well as the $\bar K N$ thresholds in the $\pi\Lambda$ distributions~\cite{ParticleDataGroup:2026mpi}. In the absence of measured invariant-mass distributions for these specific four-body final states, the results are presented as predictions. The two reactions therefore provide complementary tests of the interplay of isospin violation, coupled-channel dynamics, and four-body kinematics, with particular attention to the $a_0(980)$, $\Sigma(1430)$, and $\Lambda(1670)$ regions.


\section{Formalism}\label{sec:formalism}
\begin{figure}[htbp]
\centering
\includegraphics[width=0.98\textwidth,height=0.82\textheight,keepaspectratio]{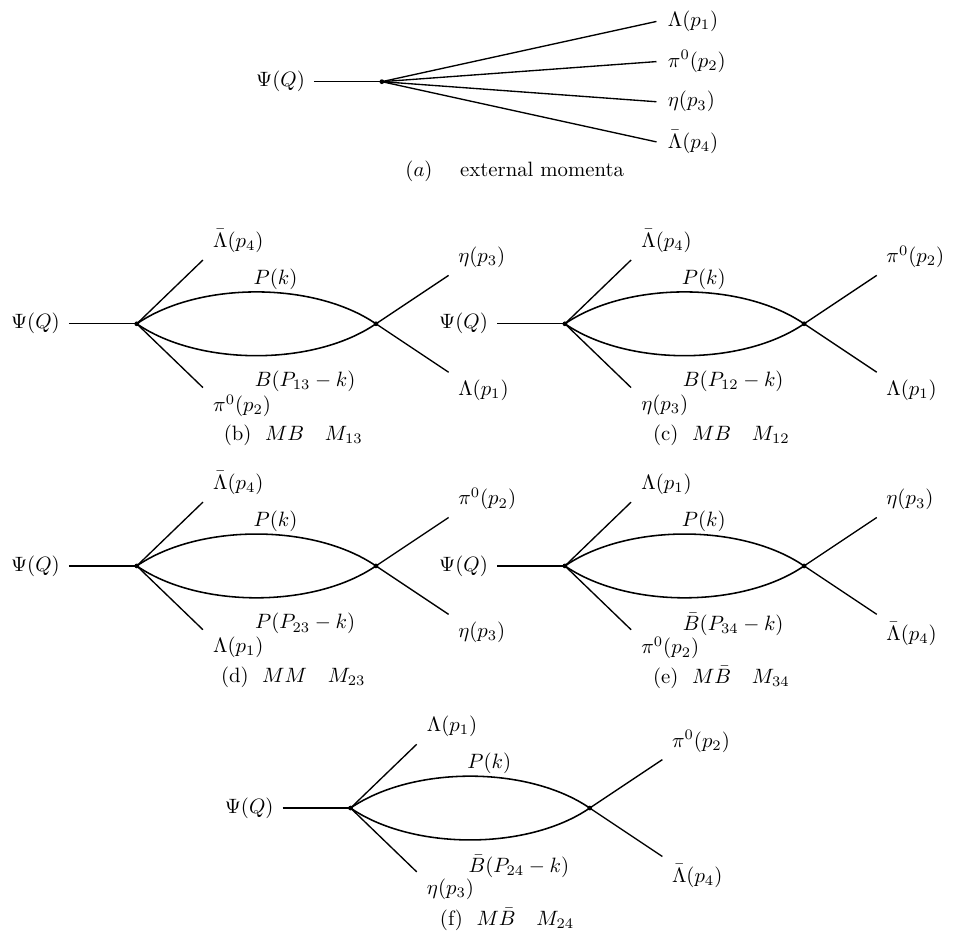}
 
\caption{Momentum assignment and organization of the  final-state interactions. Panel (a) fixes $(1,2,3,4)=(\Lambda,\pi^0,\eta,\bar\Lambda)$. Panels (b) and (c) show meson--baryon rescattering,
panel (d) meson--meson rescattering, and panels (e) and (f)
meson--antibaryon rescattering; the second spectator-meson assignment is included in the channel diagrams below.  Here $P$ denotes a pseudoscalar meson. The source vertex is a small dot; the right-hand dot represents the full coupled-channel $T$ matrix. In every loop, $P_{ij}=p_i+p_j$ and the internal momenta are $k$ and $P_{ij}-k$.}
 \label{fig:topology}
\end{figure}

We consider
\begin{equation}
 \Psi(Q)\to\Lambda(p_1)\,\pi^0(p_2)\,\eta(p_3)\,\bar\Lambda(p_4),
 \label{eq:decay}
\end{equation}
where $\Psi=J/\psi$ or $\psi(3686)$, and $Q$ and $p_i$ stand for the four-momenta of the initial and final particles. We use $M_{ij}=\sqrt{(p_i+p_j)^2}$ for the invariant masses, with the particle numbering given in Eq.~\eqref{eq:decay}.
The decay starts with the production of a baryon--antibaryon pair
and two pseudoscalar mesons from a light-flavor SU(3) singlet.
We then consider the final-state interactions in the
meson--meson, meson--baryon, and meson--antibaryon systems.
 We also consider the conversion of $\pi_3$ into $\eta$ and $\eta_I$ into $\pi^0$ through $\pi^0$--$\eta$ mixing where $\pi_3$ and $\eta_I$ stand for the neutral fields before isospin mixing. The different mechanisms are shown schematically in Fig.~\ref{fig:topology}.

We follow the usual approach of describing the interaction of each hadron pair with its coupled-channel scattering amplitude~\cite{Oset:2016lyh,Song:2022kac}. The contributions from different pairs are added coherently, while genuine three- and four-particle rescattering is not included. This approximation allows us to relate the structures in the four-body mass distributions to the two-body amplitudes entering the calculation.

\subsection{Primary production and SU(3) flavor weights}
The $c\bar c$ system has no light-flavor quantum numbers. For the primary production vertex we take the three SU(3)-invariant single-trace combinations
\begin{align}
 \mathcal H={}&\At\langle\bar BPPB\rangle\nonumber\\
 &+\Bt\langle\bar BPBP\rangle\nonumber\\
 &+\Ct\langle\bar BBPP\rangle.
 \label{eq:source}
\end{align}
where $\langle\cdots\rangle$ denotes a trace in flavor space.
As in previous related works~\cite{Ren:2026ila,Ikeno:2026zpv,Dai:2026zqn,CesarSobrinho:2026vhe}, we take the single trace terms which are dominant in the large-$N_c$ counting~\cite{Manohar:1998xv,Abreu:2023yvf}.
The meson and baryon matrices are
\begin{equation}
P=\begin{pmatrix}
\dfrac{\pi_3}{\sqrt2}+\dfrac{\eta_I}{\sqrt3}&\pi^+&K^+\\[2pt]
\pi^-&-\dfrac{\pi_3}{\sqrt2}+\dfrac{\eta_I}{\sqrt3}&K^0\\[2pt]
K^-&\bar K^0&-\dfrac{\eta_I}{\sqrt3}
\end{pmatrix},
\label{eq:P}
\end{equation}
\begin{equation}
B=\begin{pmatrix}
\dfrac{\Sigma^0}{\sqrt2}+\dfrac{\Lambda}{\sqrt6}&\Sigma^+&p\\[2pt]
\Sigma^-&-\dfrac{\Sigma^0}{\sqrt2}+\dfrac{\Lambda}{\sqrt6}&n\\[2pt]
\Xi^-&\Xi^0&-\dfrac{2\Lambda}{\sqrt6}
\end{pmatrix},
\label{eq:B}
\end{equation}
\begin{equation*}
\overline{B}=\begin{pmatrix}
\dfrac{\bar\Sigma^0}{\sqrt2}+\dfrac{\bar\Lambda}{\sqrt6}&\bar\Sigma^+&\bar\Xi^+\\[2pt]
\bar\Sigma^-&-\dfrac{\bar\Sigma^0}{\sqrt2}+\dfrac{\bar\Lambda}{\sqrt6}&\bar\Xi^0\\[2pt]
\bar p&\bar n&-\dfrac{2\bar\Lambda}{\sqrt6}
\end{pmatrix}.
\end{equation*}
The matrix $\bar B$ is the Dirac-adjoint baryon matrix. The physical $\pi^0$ and $\eta$ fields are introduced below. We use the $\eta$--$\eta'$ mixing convention of Ref.~\cite{Bramon:1992kr}, as in Refs.~\cite{Song:2022kac,Lin:2021isc}.

By expanding the SU(3) flavor structures in
Eq.~\eqref{eq:source}, we obtain the primary production
channels and their corresponding weights for the
$\Lambda\bar\Lambda\pi^0\eta$ decay.
We obtain
\begin{align}
 \mathcal H_{\Lambda\bar\Lambda}={}&\bar\Lambda\Lambda\bigg\{
 \frac{\At+\Bt+\Ct}{6}\pi_3\pi_3\nonumber\\
 &+\frac{\At+\Bt+\Ct}{3}\pi^+\pi^-\nonumber\\
 &+\frac{\At+\Bt+\Ct}{3}\eta_I\eta_I\nonumber\\
 &+\frac{5\At-4\Bt+5\Ct}{6}K^+K^-\nonumber\\
 &+\frac{5\At-4\Bt+5\Ct}{6}K^0\bar K^0\bigg\}.
 \label{eq:neutral-source}
\end{align}
Equation~\eqref{eq:neutral-source} gives the relative production weights of the $\pi\pi$, $K\bar K$, and $\eta\eta$ pairs before final-state interaction. It is important to note that there is no $\pi_3\eta_I$ term at this stage. The $\Lambda\bar\Lambda$ pair has isospin zero, while $\pi^0\eta$ has isospin one. The transition to the physical final state therefore requires isospin breaking. In our approach this comes from the charged-hadron mass differences in the intermediate loops and from $\pi^0$--$\eta$ mixing.

\subsection{Coupled-channel amplitudes}
We first consider the interactions of the two pseudoscalar mesons. For this sector, denoted by MM, we work with the six charge channels
\begin{equation}
 \pi^+\pi^-,\quad\frac{\pi^0\pi^0}{\sqrt2},\quad K^+K^-,\quad
 K^0\bar K^0,\quad\frac{\eta\eta}{\sqrt2},\quad\pi^0\eta,
 \label{eq:MMchannels}
\end{equation}
using the potential of Ref.~\cite{Lin:2021isc}, given explicitly in its Appendix~A, Eq.~(A4). The calculation employs $f=93$ MeV and a cutoff $q_{\max}=650$ MeV. For the meson--baryon sector, denoted by MB, we retain the ten charge channels
\begin{gather}
 K^-p,\ \bar K^0n,\ \pi^0\Lambda,\ \pi^0\Sigma^0,\ \eta\Lambda,\ \eta\Sigma^0,\nonumber\\
 \pi^+\Sigma^-,\ \pi^-\Sigma^+,\ K^+\Xi^-,\ K^0\Xi^0.
 \label{eq:MBchannels}
\end{gather}
The factors $1/\sqrt{2}$ in Eq.~\eqref{eq:MMchannels} account for the normalization of identical-meson states. In both sectors the two-body scattering matrix is obtained by solving the on-shell coupled-channel equation~\cite{Oller:1997ti,Oset:1997it},
\begin{equation}
 T(W)=[1-V(W)G(W)]^{-1}V(W),\qquad W=\sqrt{s},
 \label{eq:T}
\end{equation}
with $V(W)$ the transition potential and $G(W)$ the loop function or two hadron propagators given below.

\subsubsection{Meson--baryon potentials}
For the MB interaction we use the $S$-wave Weinberg--Tomozawa potential~\cite{Oset:1997it,Duan:2024okk,Li:2025yad},
\begin{align}
 V_{ij}(W)&=-\frac{C_{ij}}{4f_if_j}\,[k_i^0(W)+k_j^0(W)],\nonumber\\
 k_i^0(W)&=\frac{W^2+m_i^2-M_i^2}{2W},
 \label{eq:VMB}
\end{align}
{where $m_i$ and $M_i$ are the meson and baryon masses in channel $i$, respectively, and $f_i$ is the corresponding channel parameter. The coefficients $C_{ij}$ are obtained from Refs.~\cite{Oset:1997it,Jido:2003cb}}.

{For the $\eta\Lambda$ final-state interaction, the channel-dependent parameters $(f_i,q_{\max,i})$ are taken from Ref.~\cite{Duan:2024okk}. For the channel families $\bar K N$, $\pi\Lambda$, $\pi\Sigma$, $\eta\Lambda$, $\eta\Sigma$, and $K\Xi$, the corresponding values, in MeV, are $(42.94,400.95)$, $(40.35,860.65)$, $(179.54,1257.47)$, $(103.29,828.91)$, $(108.40,1043.17)$, and $(147.15,1389.43)$, respectively. For the $\pi^0\Lambda$ final-state interaction, we use $f_i=1.15f_\pi=106.95$ MeV, with $f_\pi=93$ MeV, and $q_{\max,i}=630$ MeV for all ten channels, following Ref.~\cite{Oset:1997it}. 
}

\subsubsection{Loop functions}
The loop functions appearing in Eq.~\eqref{eq:T} are calculated with a three-momentum cutoff. For the MM and MB channels we use, respectively~\cite{Oller:1997ti,Oset:1997it},
\begin{align}
 G^{\MM}_{i,\mathrm{cut}}(W)={}&
 \int_0^{q_{\max,i}}\frac{q^2\dd q}{2\pi^2}
 \frac{\omega_{i1}+\omega_{i2}}{2\omega_{i1}\omega_{i2}}
 \frac{1}{W^2-(\omega_{i1}+\omega_{i2})^2+i0},
 \label{eq:GMM}\\
 G^{\MB}_{i,\mathrm{cut}}(W)={}&2M_i
 \int_0^{q_{\max,i}}\frac{q^2\dd q}{2\pi^2}
 \frac{\omega_i+E_i}{2\omega_i E_i}
 \frac{1}{W^2-(\omega_i+E_i)^2+i0},
 \label{eq:GMB}
\end{align}
where $\omega_{ij}=\sqrt{m_{ij}^2+q^2}$, $\omega_i=\sqrt{m_i^2+q^2}$, and $E_i=\sqrt{M_i^2+q^2}$. The prefactor $2M_i$ in the MB loop is consistent with the normalization adopted for the baryon scattering amplitudes~\cite{MandlShaw2010}.

We use the physical hadron masses in the loop functions
to account for the threshold differences among the charge
channels. The loop functions are regularized using the cutoff
parameters specified for the corresponding interactions.

{
To avoid the region where the on-shell momentum exceeds the
corresponding three-momentum cutoff, we introduce an exponential
suppression of the rescattering contributions above a cutoff
invariant mass~\cite{Debastiani:2016ayp}. For invariant masses
above $M_{\rm cut}$, the $GT$ amplitudes are extrapolated according to
\begin{equation}
 [G_iT_{i,f}](W)=
 [G_iT_{i,f}]_{\rm cut}(M_{\rm cut}^{(f)})
 e^{-\alpha(W-M_{\rm cut}^{(f)})},
 \quad W>M_{\rm cut}^{(f)},
 \label{eq:GT-tail}
\end{equation}
where $[G_iT_{i,f}]_{\rm cut}$ denotes the amplitude evaluated
with the original three-momentum cutoff. We use
$\alpha=0.0054$~MeV$^{-1}$ and
\begin{equation}
 M_{\rm cut}^{(f)}=
 \begin{cases}
 1100~\mathrm{MeV}, & f=\pi^0\eta,\\
 1700~\mathrm{MeV}, & f=\pi^0\Lambda,\ \pi^0\bar\Lambda,\\
 1900~\mathrm{MeV}, & f=\eta\Lambda,\ \eta\bar\Lambda.
 \end{cases}
 \label{eq:GT-cutoff}
\end{equation}

}

\subsection{Final-state rescattering amplitudes}

\subsubsection{Meson--meson rescattering}
\begin{figure}[htbp]
\centering
\includegraphics[width=0.98\textwidth,height=0.82\textheight,keepaspectratio]{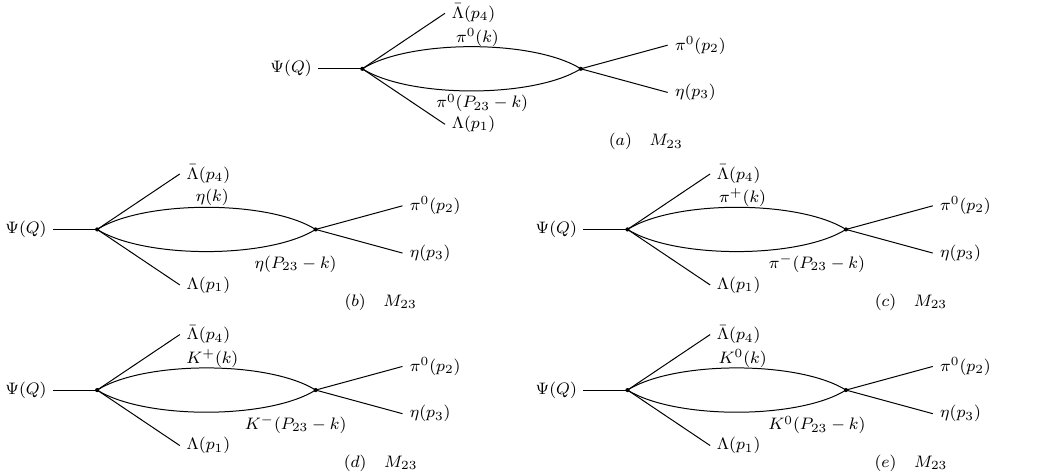}
 
\caption{The five MM contributions to Eq.~\eqref{eq:tMM}. Both baryons are spectators; the intermediate pairs are $\pi^0\pi^0$, $\eta\eta$, $\pi^+\pi^-$, $K^+K^-$, and $K^0\bar K^0$. Each pair rescatters into $\pi^0\eta$ at $M_{23}$. The identical-meson factors are contained in the coefficients of Eq.~\eqref{eq:tMM}.}
 \label{fig:mm-channels}
\end{figure}

We now obtain the decay amplitudes from the production weights and the two-body scattering matrices. We begin with the diagrams in which the two mesons interact while $\Lambda$ and $\bar\Lambda$ remain spectators. From Eq.~\eqref{eq:neutral-source}, the MM amplitude is~\cite{Lin:2021isc,Song:2022kac}
\begin{align}
 t_{\MM}(M_{23})={}&
 \frac{\At+\Bt+\Ct}{3}\,\GT{\pip}{\pieta}\nonumber\\
 &+\frac{\At+\Bt+\Ct}{3\sqrt2}\,\GT{\pin}{\pieta}\nonumber\\
 &+\frac{5\At-4\Bt+5\Ct}{6}\,\GT{\kkc}{\pieta}\nonumber\\
 &+\frac{5\At-4\Bt+5\Ct}{6}\,\GT{\kkn}{\pieta}\nonumber\\
 &+\frac{\sqrt2(\At+\Bt+\Ct)}{3}\,\GT{\eta\eta}{\pieta}.
 \label{eq:tMM}
\end{align}
Here $G_iT_{i,\pi^0\eta}$ denotes the propagation of the intermediate meson pair $i$, followed by its transition to the observed $\pi^0\eta$ pair. All functions in Eq.~\eqref{eq:tMM} are evaluated at $W=M_{23}$. The coefficients already contain the normalization of the identical-meson channels. The five possible intermediate meson pairs are shown in Fig.~\ref{fig:mm-channels}.

\subsubsection{Meson--baryon and meson--antibaryon rescattering}
\begin{figure}[htbp]
\centering
\includegraphics[width=0.98\textwidth,height=0.82\textheight,keepaspectratio]{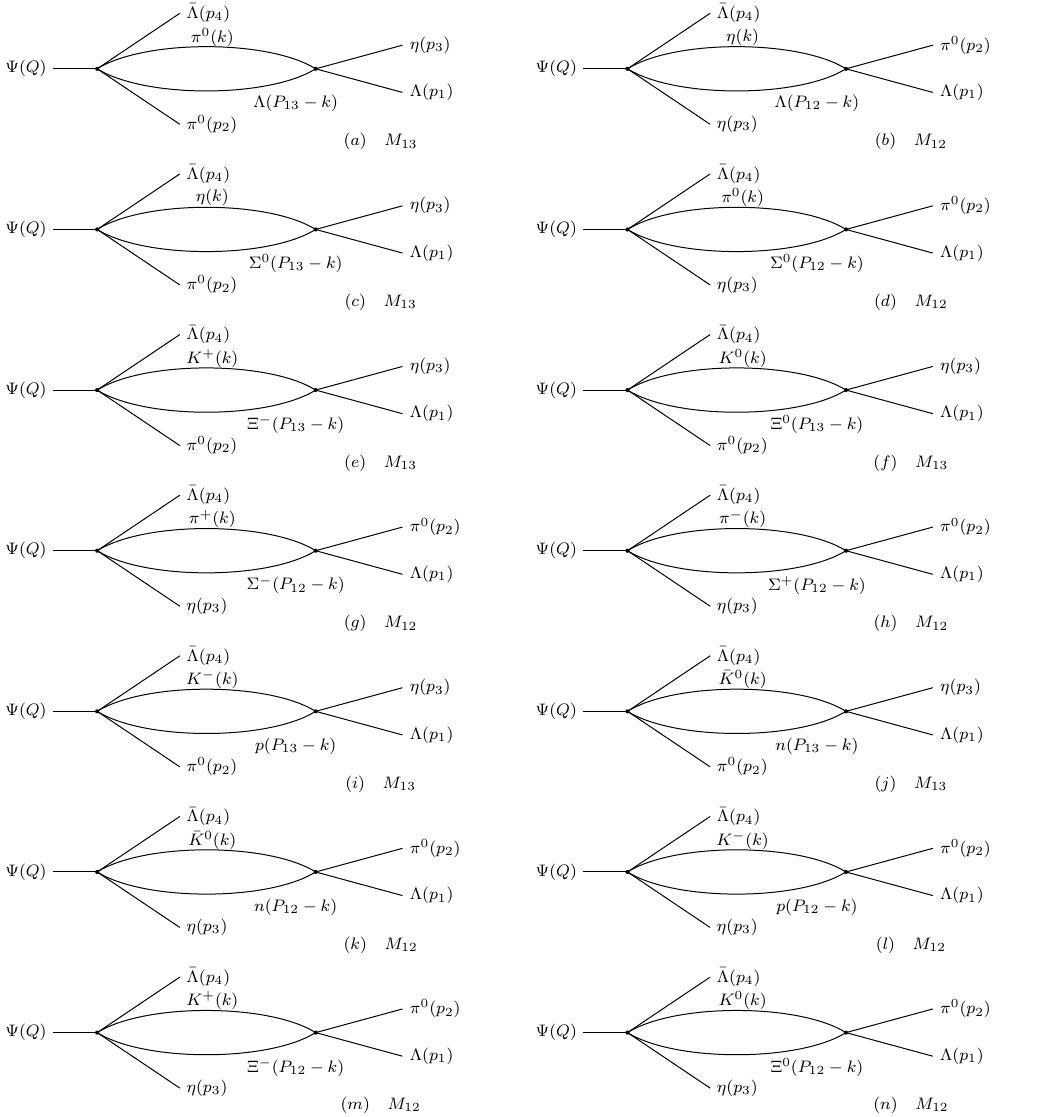}
 
\caption{All fourteen MB diagrams contributing to Eq.~\eqref{eq:tMB}, with $\bar\Lambda(p_4)$ as spectator. In (a), (c), (e), (f), (i), and (j), the spectator meson is $\pi^0(p_2)$ and the final interacting pair is $\eta\Lambda$ at $M_{13}$. The remaining eight diagrams have $\eta(p_3)$ as spectator and end in $\pi^0\Lambda$ at $M_{12}$. The two-body amplitude at the right vertex accounts for the transition scattering matrix evaluated with Eq.(~\ref{eq:T}) with all the coupled channels. }
 \label{fig:mb-channels}
\end{figure}

We consider next the interaction of one pseudoscalar meson with the $\Lambda$, leaving the $\bar\Lambda$ as spectator. There are two possibilities: the $\pi^0$ is the other spectator and the $\eta\Lambda$ pair interacts at $M_{13}$, or the $\eta$ is the spectator and the $\pi^0\Lambda$ pair interacts at $M_{12}$. The corresponding production and rescattering terms give 
\begingroup\allowdisplaybreaks[0]
\begin{align}
 t_{\MB}={}&\bigg[
 -\frac{2\At-\Bt-\Ct}{2\sqrt3}\,\MBGT{\kp}{\eL}\nonumber\\
 &+\frac{2\At-\Bt-\Ct}{2\sqrt3}\,\MBGT{\kn}{\eL}\nonumber\\
 &+\frac{\At+\Bt+\Ct}{3}\,\MBGT{\pL}{\eL}\nonumber\\
 &+\frac{\sqrt2(\At+\Bt+\Ct)}{3}\,\MBGT{\eS}{\eL}\nonumber\\
 &+\frac{\At+\Bt-2\Ct}{2\sqrt3}\,\MBGT{\kx}{\eL}\nonumber\\
 &-\frac{\At+\Bt-2\Ct}{2\sqrt3}\,\MBGT{\kz}{\eL}
 \bigg]_{M_{13}}\nonumber\\[2pt]
 &+\bigg[
 \frac{\Bt}{\sqrt2}\,\MBGT{\kp}{\pL}\nonumber\\
 &+\frac{\Bt}{\sqrt2}\,\MBGT{\kn}{\pL}\nonumber\\
 &+\frac{\sqrt2(\At+\Bt+\Ct)}{3}\,\MBGT{\pS}{\pL}\nonumber\\
 &+\frac{2(\At+\Bt+\Ct)}{3}\,\MBGT{\eL}{\pL}\nonumber\\
 &+\frac{\sqrt2(\At+\Bt+\Ct)}{3}\,\MBGT{\psm}{\pL}\nonumber\\
 &+\frac{\sqrt2(\At+\Bt+\Ct)}{3}\,\MBGT{\msp}{\pL}\nonumber\\
 &+\frac{\Bt}{\sqrt2}\,\MBGT{\kx}{\pL}\nonumber\\
 &+\frac{\Bt}{\sqrt2}\,\MBGT{\kz}{\pL}
 \bigg]_{M_{12}}.
 \label{eq:tMB}
\end{align}
\endgroup
The first bracket contains six intermediate channels that end in $\eta\Lambda$ at $M_{13}$; the second contains eight channels that end in $\pi^0\Lambda$ at $M_{12}$. The latter includes the $I=1$ interaction around the $\bar K N$ thresholds, while the $\eta\Lambda$ amplitude contains the isoscalar dynamics associated with the $\Lambda(1670)$. We show the fourteen diagrams in Fig.~\ref{fig:mb-channels}.

\begin{figure}[htbp]
\centering
\includegraphics[width=0.98\textwidth,height=0.82\textheight,keepaspectratio]{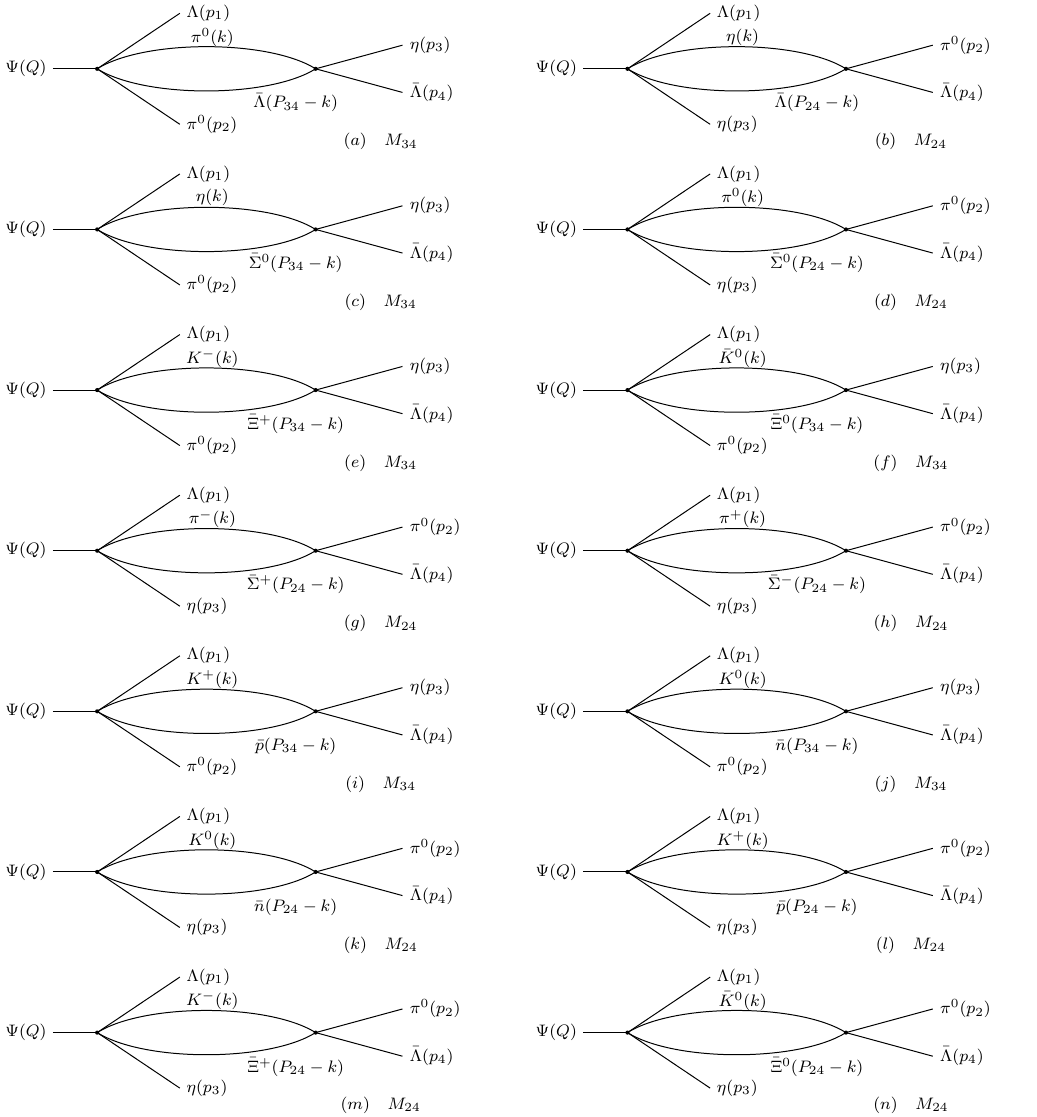}
 
\caption{Charge-conjugated diagrams for Eq.~\eqref{eq:tMbarB}. The spectator baryon is $\Lambda(p_1)$. Panels (a), (c), (e), (f), (i), and (j) contribute at $M_{34}$; the other eight contribute at $M_{24}$. The superscript on a charged antibaryon denotes its electric charge, e.g.\ $\bar\Sigma^+$ is the antiparticle of $\Sigma^-$. The intermediate channels and source phases are charge conjugated consistently with the MB diagrams in Fig.~\ref{fig:mb-channels}.}
 \label{fig:mbarb-channels}
\end{figure}

The meson--antibaryon amplitude is constructed in the same way, with $\Lambda$ as spectator. Charge conjugation relates the strong amplitudes and the source coefficients to those just obtained. The result has the form
\begin{equation}
 t_{\MbarB}(M_{24},M_{34})=
 \left.t_{\MB}(M_{12},M_{13})\right|_{M_{12}\to M_{24},\,M_{13}\to M_{34}}.
 \label{eq:tMbarB}
\end{equation}
Here the charged intermediate states are replaced by their antiparticles, with the same phase convention. The associated diagrams are shown in Fig.~\ref{fig:mbarb-channels}.

\subsection{\texorpdfstring{Isospin violation from $\pi^0$--$\eta$ mixing}{Isospin violation from pi0-eta mixing}}
\subsubsection{Mixing-induced tree contribution}
\begin{figure}[htbp]
\centering
\includegraphics[width=0.98\textwidth,height=0.82\textheight,keepaspectratio]{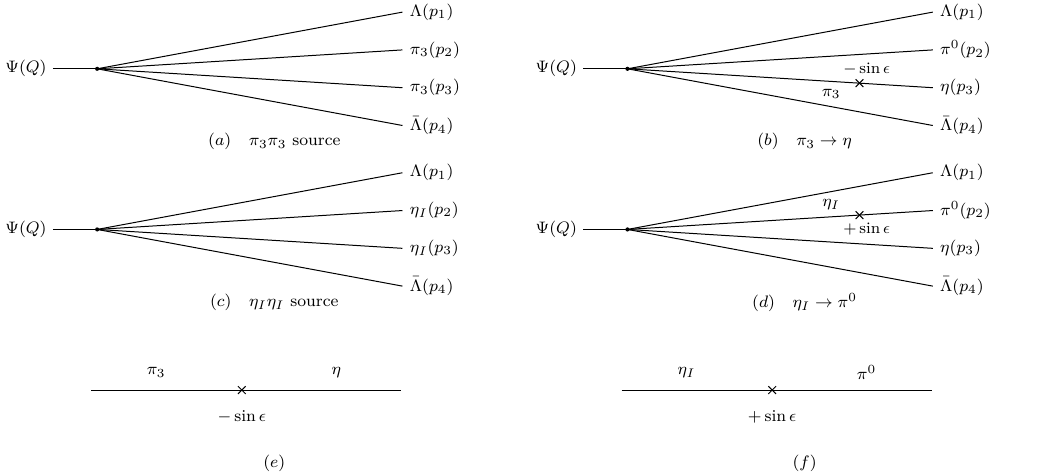}
 
\caption{External mixing at tree level. Panels (a) and (c) show the isospin-conserving $\pi_3\pi_3$ and $\eta_I\eta_I$ sources. In (b), the field on the $p_3$ leg is converted as $\pi_3\to\eta$, with factor $-\sin\epsilon$; in (d), the $p_2$ leg is converted as $\eta_I\to\pi^0$, with factor $+\sin\epsilon$. Panels (e) and (f) summarize these external-leg rules. The cross marks the external conversion. The coefficients in Eq.~\eqref{eq:tree} include the two choices of the converted identical field.}
 \label{fig:tree-mixing}
\end{figure}

Besides the mass differences in the loops, we consider the mixing of the neutral fields $\pi_3$ and $\eta_I$. We write the physical fields as
\begin{align}
 \pi^0&=\cos\epsilon\,\pi_3+\sin\epsilon\,\eta_I,\nonumber\\
 \eta&=-\sin\epsilon\,\pi_3+\cos\epsilon\,\eta_I.
 \label{eq:mixing}
\end{align}
With this convention, $\pi_3\to\eta$ gives $-\sin\epsilon$, while $\eta_I\to\pi^0$ gives $+\sin\epsilon$. 
Keeping terms to first order in the small mixing angle,
the two meson sources in Eq.~\eqref{eq:neutral-source}
give the direct $\pi^0\eta$ amplitude
\begin{equation}
\begin{aligned}
 t_{\Tree}
 &=-\frac{2(\At+\Bt+\Ct)}{6}\sin\epsilon\\
&\quad+\frac{2(\At+\Bt+\Ct)}{3}\sin\epsilon\\
 &=\frac{\At+\Bt+\Ct}{3}\sin\epsilon.
\end{aligned}
\label{eq:tree}
\end{equation}
The first term comes from the $\pi_3\pi_3$ source and the second from $\eta_I\eta_I$. The factor of two takes into account the two equivalent neutral fields before conversion. The diagrams are shown in Fig.~\ref{fig:tree-mixing}.

\subsubsection{Meson--meson mixing contribution}

The same mixing also acts on the meson pairs after rescattering.
For the $\pi_3\pi_3$ and $\eta_I\eta_I$ channels, we obtain
\begingroup\allowdisplaybreaks[0]
\begin{align}
 t_{\mix,\MM}={}&\sin\epsilon\bigg[
 \frac{2\sqrt2(\At+\Bt+\Ct)}{3}\,\GT{\pip}{\eta\eta}\nonumber\\
 &-\frac{2\sqrt2(\At+\Bt+\Ct)}{3}\,\GT{\pip}{\pin}\nonumber\\
 &+\frac{2(\At+\Bt+\Ct)}{3}\,\GT{\pin}{\eta\eta}\nonumber\\
 &-\frac{2(\At+\Bt+\Ct)}{3}\,\GT{\pin}{\pin}\nonumber\\
 &+\frac{\sqrt2(5\At-4\Bt+5\Ct)}{3}\,\GT{\kkc}{\eta\eta}\nonumber\\
 &-\frac{\sqrt2(5\At-4\Bt+5\Ct)}{3}\,\GT{\kkc}{\pin}\nonumber\\
 &+\frac{\sqrt2(5\At-4\Bt+5\Ct)}{3}\,\GT{\kkn}{\eta\eta}\nonumber\\
 &-\frac{\sqrt2(5\At-4\Bt+5\Ct)}{3}\,\GT{\kkn}{\pin}\nonumber\\
 &+\frac{4(\At+\Bt+\Ct)}{3}\,\GT{\eta\eta}{\eta\eta}\nonumber\\
 &-\frac{4(\At+\Bt+\Ct)}{3}\,\GT{\eta\eta}{\pin}\bigg]_{M_{23}}.
 \label{eq:mixMM}
\end{align}
\endgroup

\begin{figure}[htbp]
\centering
\includegraphics[width=0.98\textwidth,height=0.82\textheight,keepaspectratio]{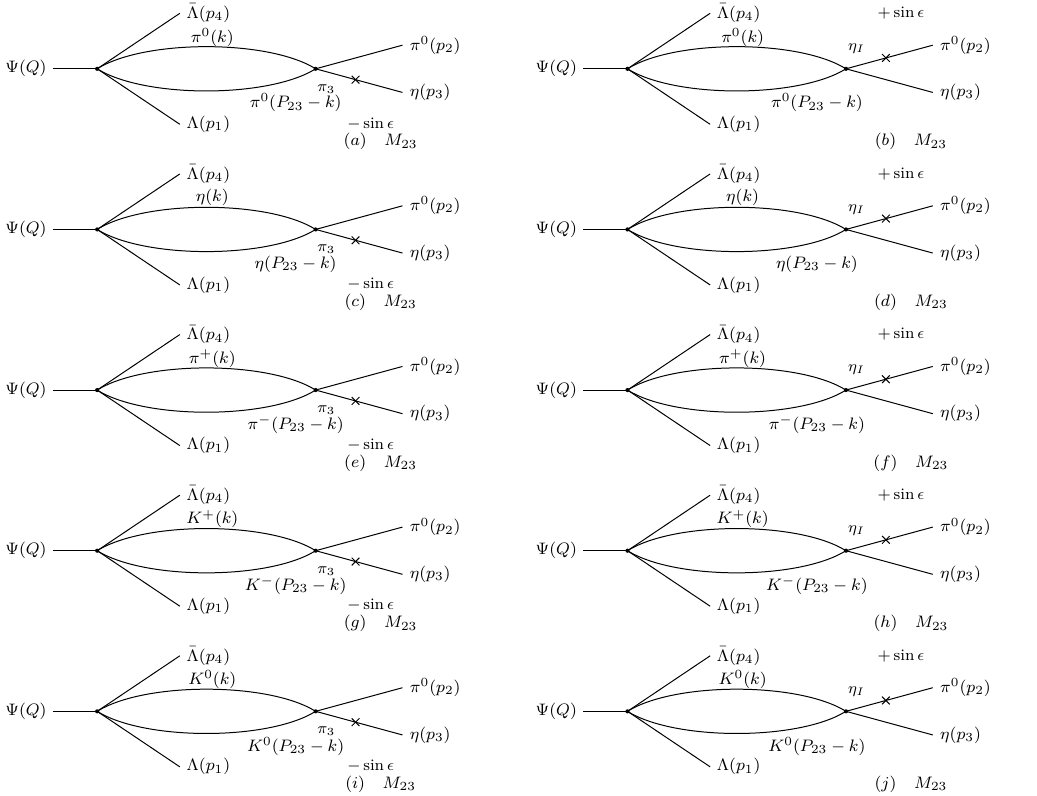}
 \caption{Diagrams for the MM mixing contributions in
Eq.~\eqref{eq:mixMM}. In each row, the intermediate meson pair
is the same. The left column ends in a $\pi_3\pi_3$ pair
followed by $\pi_3(p_3)\to\eta(p_3)$, while the right column
ends in $\eta_I\eta_I$ followed by
$\eta_I(p_2)\to\pi^0(p_2)$. All loops are evaluated at
$M_{23}$. The coefficients in Eq.~\eqref{eq:mixMM} include
the normalization and mixing-diagram factors.}
 \label{fig:mm-mixing}
\end{figure}

All terms in Eq.~\eqref{eq:mixMM} are evaluated at $M_{23}$.
Each of the five intermediate meson pairs can rescatter into
$\eta_I\eta_I$ or $\pi_3\pi_3$. The corresponding mixing
contributions have opposite signs.
The factors associated with identical mesons and the two possible external conversions are included explicitly. These ten diagrams are collected in Fig.~\ref{fig:mm-mixing}.

\subsubsection{Meson--baryon and meson--antibaryon mixing contributions}

\begin{figure}[htbp]
\centering
\includegraphics[width=0.98\textwidth,height=0.82\textheight,keepaspectratio]{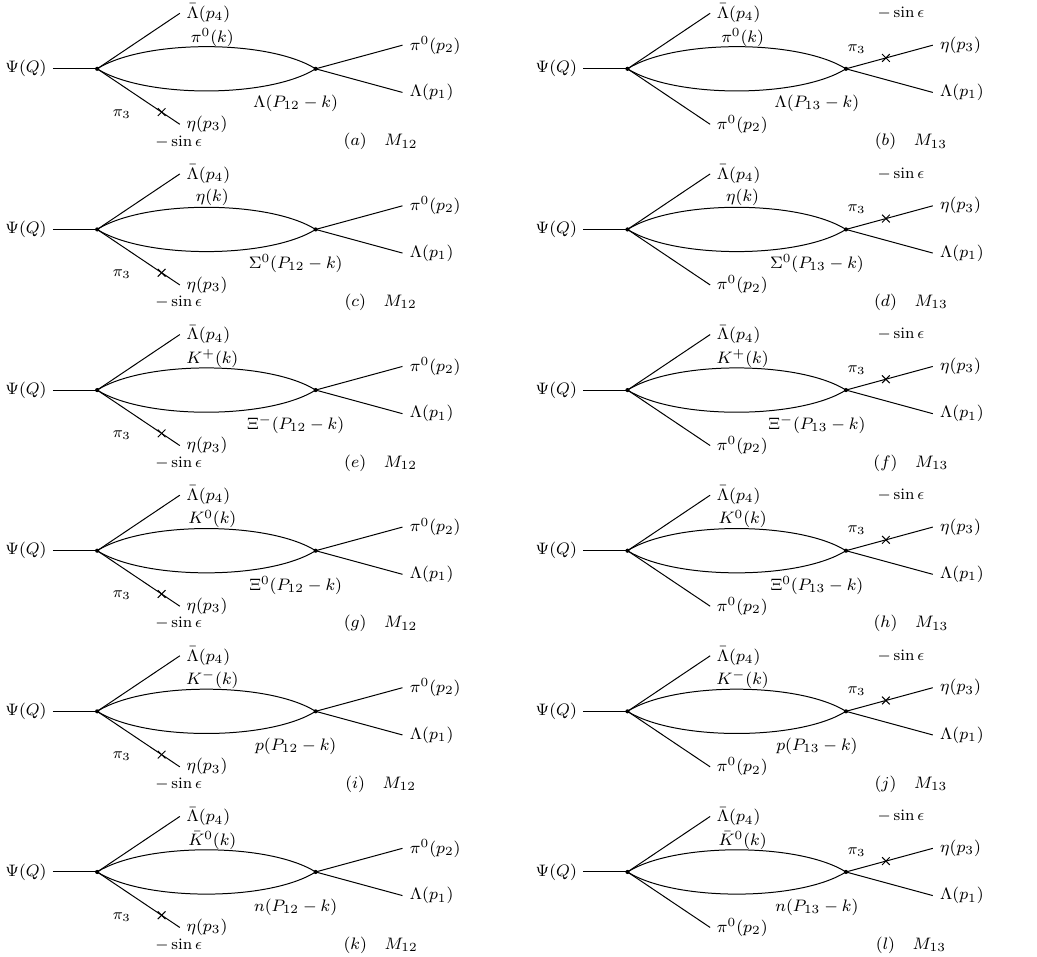}
 
\caption{The twelve MB mixing diagrams with an auxiliary $\pi_3\pi_3$ state. The six intermediate channels are $\pi^0\Lambda$, $\eta\Sigma^0$, $K^+\Xi^-$, $K^0\Xi^0$, $K^-p$, and $\bar K^0n$. In the left column, $\pi_3(p_3)\to\eta(p_3)$ occurs on the spectator leg and the scattering pair has mass $M_{12}$. In the right column, the same conversion occurs after rescattering, with pair mass $M_{13}$. Each cross contributes $-\sin\epsilon$; the channel-dependent signs and coefficients are those of Eq.~\eqref{eq:mixMB}.}
 \label{fig:mb-mix-pi}
\end{figure}

\begin{figure}[htbp]
\centering
\includegraphics[width=0.98\textwidth,height=0.82\textheight,keepaspectratio]{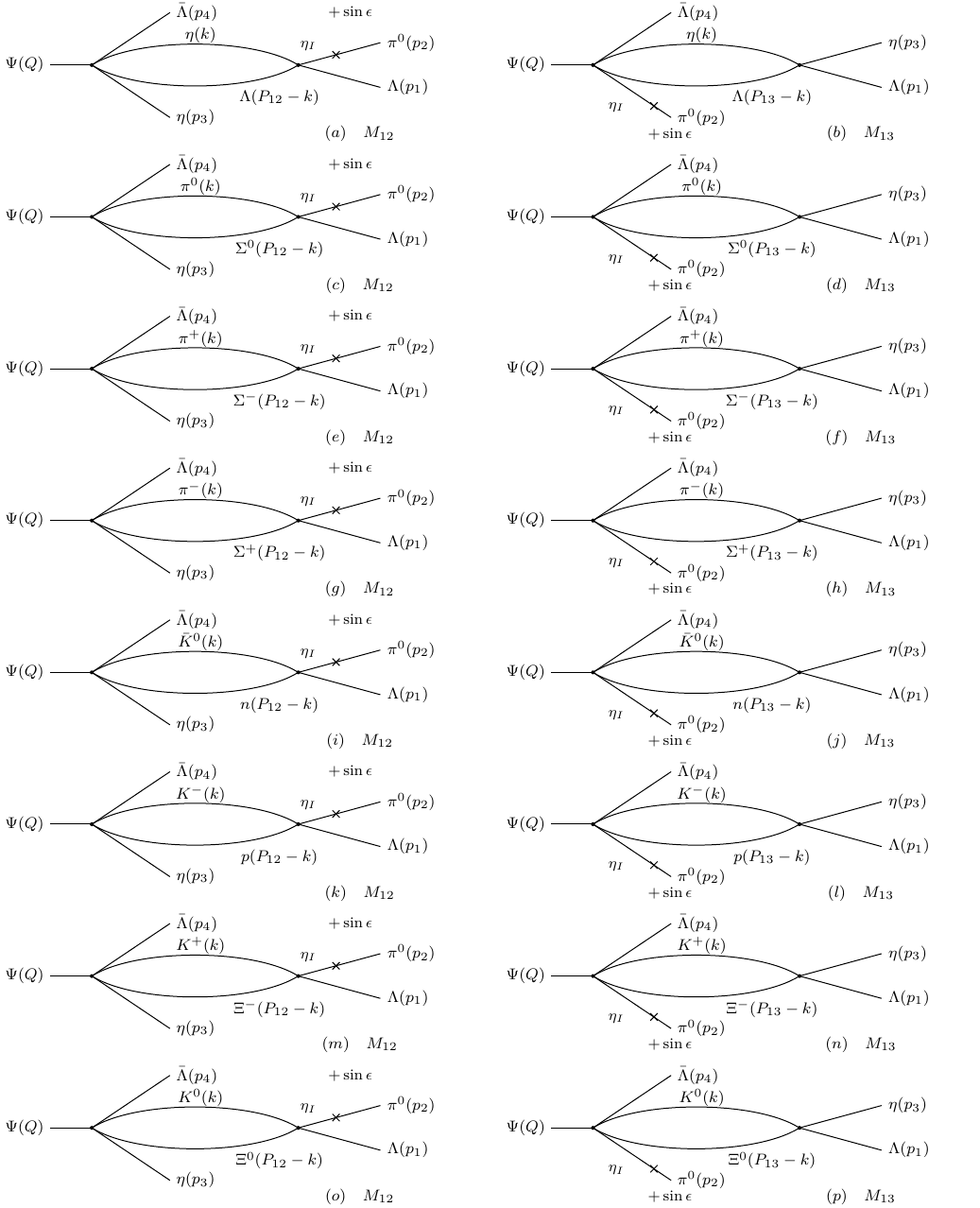}
 
\caption{The sixteen MB mixing diagrams with an auxiliary $\eta_I\eta_I$ state. Each of the eight intermediate channels contributes at both $M_{12}$ (left) and $M_{13}$ (right). The conversion $\eta_I(p_2)\to\pi^0(p_2)$ occurs on the rescattered leg on the left and on the spectator leg on the right, and contributes $+\sin\epsilon$. The first row contains the explicit $\eta\Lambda$ elastic amplitude in Eq.~\eqref{eq:mixMB}.}
 \label{fig:mb-mix-eta}
\end{figure}

Finally, we consider mixing together with MB rescattering.
The mixing contributes through both the spectator meson and
the meson--baryon final-state interaction. Hence, the
$\pi^0\Lambda$ and $\eta\Lambda$ amplitudes contribute at
both $M_{12}$ and $M_{13}$. Using the same production weights
as in Eq.~\eqref{eq:tMB}, we find
\begingroup\allowdisplaybreaks[0]
\begin{align}
t_{\mix,\MB}={}&
\sin\epsilon\bigg[
 \frac{2(\At+\Bt+\Ct)}{3}\,\MBGT{\eL}{\eL}
 \nonumber\\
&+\frac{\sqrt2(\At+\Bt+\Ct)}{3}\,\MBGT{\pS}{\eL}
 \nonumber\\
&+\frac{\sqrt2(\At+\Bt+\Ct)}{3}\,\MBGT{\psm}{\eL}
 \nonumber\\
&+\frac{\sqrt2(\At+\Bt+\Ct)}{3}\,\MBGT{\msp}{\eL}
 \nonumber\\
&+\frac{\Bt}{\sqrt2}\,\MBGT{\kp}{\eL}
 \nonumber\\
&+\frac{\Bt}{\sqrt2}\,\MBGT{\kn}{\eL}
 \nonumber\\
&+\frac{\Bt}{\sqrt2}\,\MBGT{\kx}{\eL}
 \nonumber\\
&+\frac{\Bt}{\sqrt2}\,\MBGT{\kz}{\eL}
 \nonumber\\
&-\frac{\At+\Bt+\Ct}{3}\,\MBGT{\pL}{\pL}
 \nonumber\\
&-\frac{\sqrt2(\At+\Bt+\Ct)}{3}\,\MBGT{\eS}{\pL}
 \nonumber\\
&+\frac{2\At-\Bt-\Ct}{2\sqrt3}\,\MBGT{\kp}{\pL}
 \nonumber\\
&-\frac{2\At-\Bt-\Ct}{2\sqrt3}\,\MBGT{\kn}{\pL}
 \nonumber\\
&-\frac{\At+\Bt-2\Ct}{2\sqrt3}\,\MBGT{\kx}{\pL}
 \nonumber\\
&+\frac{\At+\Bt-2\Ct}{2\sqrt3}\,\MBGT{\kz}{\pL}
\bigg]_{M_{12}}
\nonumber\displaybreak[1]\\[4pt]
&+\sin\epsilon\bigg[
 \frac{2(\At+\Bt+\Ct)}{3}\,\MBGT{\eL}{\eL}
 \nonumber\\
&+\frac{\sqrt2(\At+\Bt+\Ct)}{3}\,\MBGT{\pS}{\eL}
 \nonumber\\
&+\frac{\sqrt2(\At+\Bt+\Ct)}{3}\,\MBGT{\psm}{\eL}
 \nonumber\\
&+\frac{\sqrt2(\At+\Bt+\Ct)}{3}\,\MBGT{\msp}{\eL}
 \nonumber\\
&+\frac{\Bt}{\sqrt2}\,\MBGT{\kp}{\eL}
 \nonumber\\
&+\frac{\Bt}{\sqrt2}\,\MBGT{\kn}{\eL}
 \nonumber\\
&+\frac{\Bt}{\sqrt2}\,\MBGT{\kx}{\eL}
 \nonumber\\
&+\frac{\Bt}{\sqrt2}\,\MBGT{\kz}{\eL}
 \nonumber\\
&-\frac{\At+\Bt+\Ct}{3}\,\MBGT{\pL}{\pL}
 \nonumber\\
&-\frac{\sqrt2(\At+\Bt+\Ct)}{3}\,\MBGT{\eS}{\pL}
 \nonumber\\
&+\frac{2\At-\Bt-\Ct}{2\sqrt3}\,\MBGT{\kp}{\pL}
 \nonumber\\
&-\frac{2\At-\Bt-\Ct}{2\sqrt3}\,\MBGT{\kn}{\pL}
 \nonumber\\
&-\frac{\At+\Bt-2\Ct}{2\sqrt3}\,\MBGT{\kx}{\pL}
 \nonumber\\
&+\frac{\At+\Bt-2\Ct}{2\sqrt3}\,\MBGT{\kz}{\pL}
\bigg]_{M_{13}}.
\label{eq:mixMB}
\end{align}
\endgroup
The eight terms ending in $\eta\Lambda$ arise from the $\eta_I\eta_I$ source and enter with $+\sin\epsilon$; the six terms ending in $\pi^0\Lambda$ come from the $\pi_3\pi_3$ source and carry the opposite sign. In each case the two possible spectator assignments are retained. 
The corresponding diagrams are shown in
Figs.~\ref{fig:mb-mix-pi} and \ref{fig:mb-mix-eta}.

\begin{figure}[htbp]
\centering
\includegraphics[width=0.98\textwidth,height=0.82\textheight,keepaspectratio]{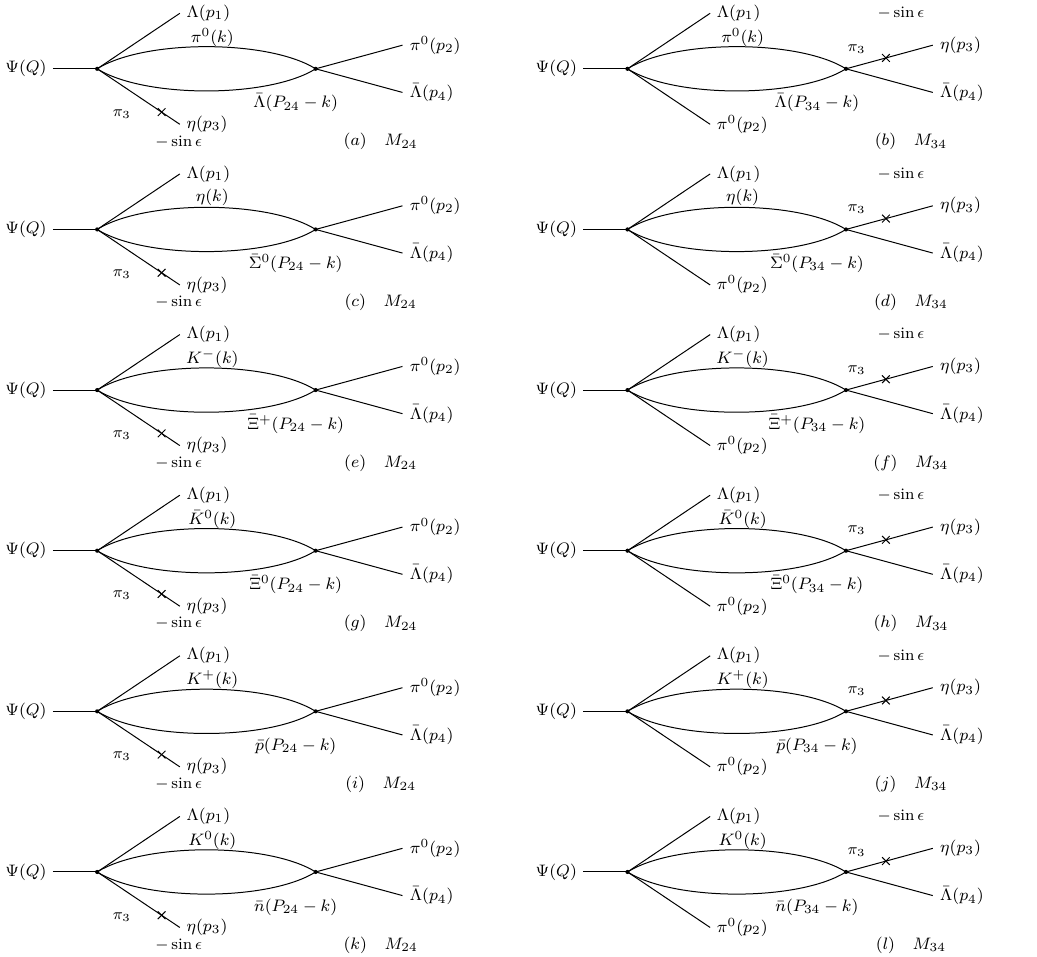}
 
\caption{Charge-conjugated $\pi_3\pi_3$ mixing diagrams for Eq.~\eqref{eq:mixMbarB}. The spectator baryon is $\Lambda(p_1)$. The left and right columns are evaluated at $M_{24}$ and $M_{34}$, respectively. The $\pi_3\to\eta$ conversion remains on the meson leg and carries $-\sin\epsilon$, with the same source coefficients as in the charge-conjugated MB expression.}
 \label{fig:mbarb-mix-pi}
\end{figure}

\begin{figure}[htbp]
\centering
\includegraphics[width=0.98\textwidth,height=0.82\textheight,keepaspectratio]{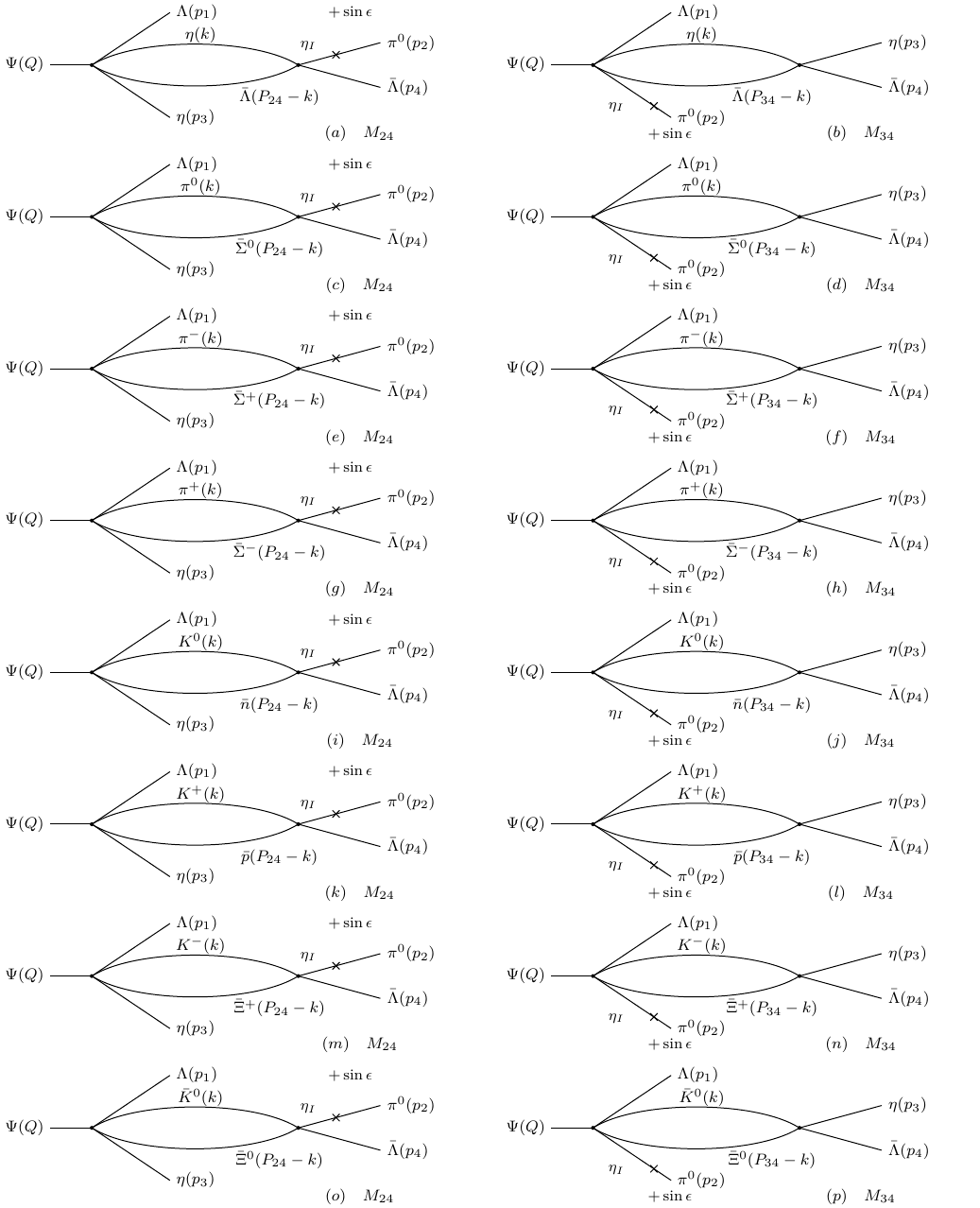}
 \caption{Diagrams for the $\eta_I\eta_I$ mixing contributions
in Eq.~\eqref{eq:mixMbarB}. The left and right columns
correspond to $M_{24}$ and $M_{34}$, respectively.
The cross represents the $\eta_I\to\pi^0$ conversion
with a factor $+\sin\epsilon$.}

 \label{fig:mbarb-mix-eta}
\end{figure}

Charge conjugation gives
\begin{equation}
 t_{\mix,\MbarB}(M_{24},M_{34})=
 \left.t_{\mix,\MB}\right|_{M_{12}\to M_{24},\,M_{13}\to M_{34}}.
 \label{eq:mixMbarB}
\end{equation} 
The diagrams for the corresponding meson--antibaryon
contributions are shown in
Figs.~\ref{fig:mbarb-mix-pi} and \ref{fig:mbarb-mix-eta}.

\subsubsection{Coherent total amplitude}

The total decay amplitude is given by
\begin{align}
 t_{\mathrm{total}}={}&t_{\MM}+t_{\MB}+t_{\MbarB}+t_{\Tree}\nonumber\\
 &+t_{\mix,\MM}+t_{\mix,\MB}+t_{\mix,\MbarB}.
 \label{eq:total}
\end{align}

In Eq.~\eqref{eq:total}, each contribution is evaluated at
the invariant mass of the corresponding interacting pair.
The different contributions are added coherently to obtain
the total decay amplitude.

\subsection{Decay width and invariant-mass distributions}


The initial charmonium state has spin one, while the $\Lambda$ and
$\bar\Lambda$ have spin $1/2$. 
The vertex structure for $\Psi(Q)\to \bar B (p')PPB(p)$ rescattering parity is given by 
\begin{equation}
 \mathcal{L} \sim \bar u(p)\epsilon_\mu\gamma^\mu\nu(p'),
 \label{eq:spin_1}
\end{equation}
which to lowest order in momentum in the $\Psi(Q)$ rest frame gives rise to the transition matrix
\begin{equation}
 t \sim \epsilon^i \langle \chi_r|\sigma^i| \chi'_r \rangle,
 \label{eq:spin_2}
\end{equation}
where $ \chi_r,~ \chi'_r$ are the Pauli bispinors of the $B$ and $\bar B$ baryons respectively.
Summing over
the final spins and averaging over the three initial polarizations,
we obtain
\begin{equation}
 \frac{1}{3}\sum_{\lambda,r,s}
 |\widetilde t_{rs}^{(\lambda)}|^2
 =\frac{1}{3}\delta_{ij}
   \operatorname{Tr}(\sigma_i\sigma_j)
   |t_{\mathrm{total}}|^2
 =2|t_{\mathrm{total}}|^2 .
 \label{eq:spin-sum}
\end{equation}
Here we have used
$\sum_\lambda\epsilon_i^{(\lambda)}
\epsilon_j^{(\lambda)}=\delta_{ij}$ in the charmonium rest frame.

The decay width is evaluated using the Lorentz-invariant four-body
phase space,
\begin{equation}
 \Gamma_\Psi=
 \frac{(2m_\Lambda)^2}{2M_\Psi}
 \int \dd\Phi_4\,2|t_{\mathrm{total}}|^2,
 \label{eq:width}
\end{equation}
where $M_\Psi$ is the mass of the initial charmonium and
\begin{equation}
 \dd\Phi_4=(2\pi)^4
 \delta^{(4)}\!\left(Q-\sum_{i=1}^4p_i\right)
 \prod_{i=1}^4
 \frac{\dd^3\bm p_i}{(2\pi)^3\,2E_i}.
 \label{eq:phase-space}
\end{equation}
The factor $(2m_\Lambda)^2$ accounts for the normalization
of the baryon spinors. We present the details on the evaluation of $ \Gamma_\Psi$  in the Appendix.

\section{Results and discussion}\label{sec:results}

We now discuss the invariant-mass distributions of
$J/\psi,\psi(3686)\to\Lambda\bar\Lambda\pi^0\eta$. With the
momentum assignment $(1,2,3,4)=(\Lambda,\pi^0,\eta,\bar\Lambda)$,
the six two-particle invariant masses are
$M_{12}=M_{\pi^0\Lambda}$, $M_{13}=M_{\eta\Lambda}$,
$M_{14}=M_{\Lambda\bar\Lambda}$, $M_{23}=M_{\pi^0\eta}$,
$M_{24}=M_{\pi^0\bar\Lambda}$, and $M_{34}=M_{\eta\bar\Lambda}$.
They are calculated from the coherent amplitude in
Eq.~\eqref{eq:total}, using the four-body phase-space integration
of Eq.~\eqref{eq:width}. The full distributions and the
corresponding phase-space distributions are normalized separately
to unit area.

\subsection{Numerical input}

The real production coefficients and the $\pi^0$--$\eta$ mixing
parameter are taken as~\cite{CesarSobrinho:2026vhe}
\begin{equation}
 \widetilde B=1,\qquad
 \frac{\widetilde A}{\widetilde B}=0.255,\qquad
 \frac{\widetilde C}{\widetilde B}=0.891,\qquad
 \sin\epsilon=0.01.
 \label{eq:parameters}
\end{equation}
The MM amplitudes are taken from Ref.~\cite{Lin:2021isc}, with
$f=93$ MeV and $q_{\max}=650$ MeV. The MB amplitudes used for the
spectra shown below are evaluated in the ten-channel charge basis
of Eq.~\eqref{eq:MBchannels}, with the interaction parameters
specified in Refs.~\cite{Duan:2024okk,Li:2024tvo,Li:2025yad}.
For the isovector interaction, we use $f=93$ MeV and
$q_{\max}=630$ MeV. The charged-channel coefficients are listed in
Table~\ref{tab:sigma-charge}. The same production parameters and
two-body input are used for the two charmonium decays.
{To study the effect of $\pi^0$--$\eta$ mixing,
we also calculate the invariant-mass distributions with
$\sin\epsilon=0$, with the same coupled-channel amplitudes
and physical hadron masses.}

\subsection{\texorpdfstring{$J/\psi\to\Lambda\bar\Lambda\pi^0\eta$}{The J/psi decay}}

\begin{figure}[htbp]
 \centering
\includegraphics[width=0.96\textwidth]{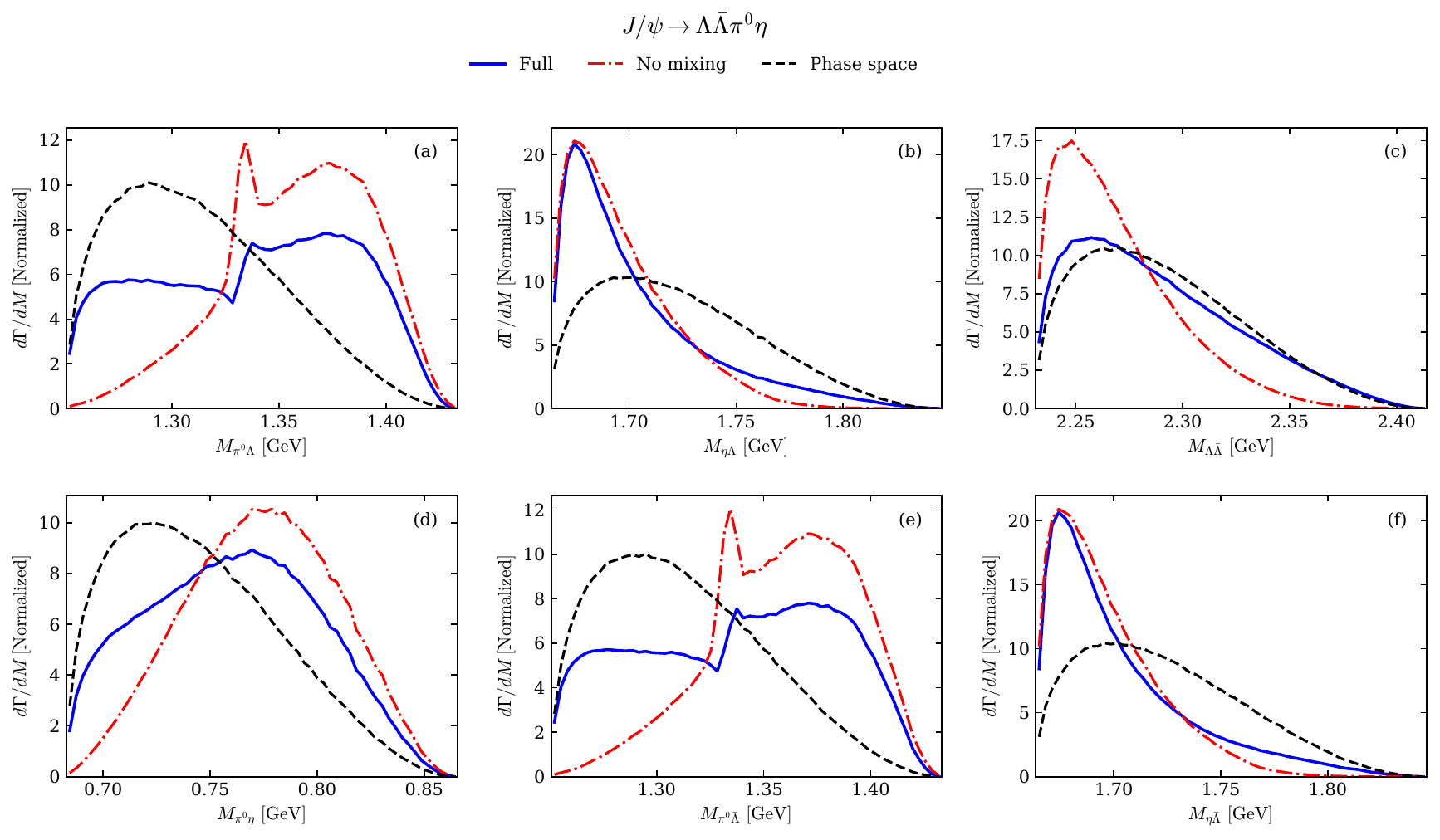}
 \caption{Normalized invariant-mass distributions for
 $J/\psi\to\Lambda\bar\Lambda\pi^0\eta$. Upper panels, from left
 to right: $M_{\pi^0\Lambda}$, $M_{\eta\Lambda}$, and
 $M_{\Lambda\bar\Lambda}$. Lower panels:
 $M_{\pi^0\eta}$, $M_{\pi^0\bar\Lambda}$, and
 $M_{\eta\bar\Lambda}$. The solid curves show the full coherent
 calculation, and the dashed curves show phase space. {The red dash-dotted curves show the results
obtained with $\sin\epsilon=0$.} Each curve
 is normalized to unit area. }
 \label{fig:jpsi}
\end{figure}

Figure~\ref{fig:jpsi} shows the six invariant-mass distributions
for $J/\psi\to\Lambda\bar\Lambda\pi^0\eta$. The limited phase space
restricts the accessible invariant-mass regions. In particular, the
maximum $\pi^0\eta$ invariant mass is
\begin{equation}
 M_{\pi^0\eta}^{\max}=M_{J/\psi}-2m_\Lambda
 =865.5\ \mathrm{MeV}.
 \label{eq:jpsi-limit}
\end{equation}
The $M_{\pi^0\eta}$ distribution therefore lies below the
$a_0(980)$ and $K\bar K$ threshold region, and shows no particular structure. The mass distribution follows basically the phase space. It exhibits a broad
maximum around $730$ MeV and varies smoothly over the
accessible range.

The upper limit of the $\pi^0\Lambda$ invariant mass is
\begin{equation}
 M_{\pi^0\Lambda}^{\max}
 =M_{J/\psi}-m_\Lambda-m_\eta
 =1433.4\ \mathrm{MeV}.
 \label{eq:piLambda-Jpsi-bound}
\end{equation}
The $K^-p$ and $\bar K^0n$ thresholds are at $1431.9$ MeV and
$1437.2$ MeV, respectively. Thus, the physical region reaches
the $K^-p$ threshold only close to the upper threshold, whereas the
$\bar K^0n$ threshold lies above it. 
The $\pi^0\Lambda$ and
$\pi^0\bar\Lambda$ distributions consequently show broad
low-mass maxima, together with a mild variation around
$1330$--$1340$ MeV, in the region of the coupled $\pi\Sigma$
thresholds.
The $\pi^0\Lambda$ channel in $I=1$ cannot generate either of the two $\Lambda(1405)$ states in the upper tail of the $\pi^0\Lambda$ distribution. However, the $\pi^0-\eta$ mixing includes amplitudes $MB\to \eta\Lambda(\to \pi^0\Lambda)$ (see diagrams to the left in Fig.~\ref{fig:mb-mix-eta}). But the $\eta\Lambda$ couples only strongly to the $\Lambda(1670)$ which is out of the range of the $\pi^0\Lambda$ invariant mass.

The clearest final-state-interaction effect appears in the
$\eta\Lambda$ and $\eta\bar\Lambda$ invariant-mass distributions.
The $\eta\Lambda$ threshold is
$m_\eta+m_\Lambda\simeq1664$ MeV, close to the
$\Lambda(1670)$ region. Both distributions rise rapidly above
threshold, reach maxima around $1680$ MeV, and then decrease
gradually. This behavior reflects the strong $\eta\Lambda$
interaction in the isoscalar coupled-channel amplitude
of Ref.~\cite{Duan:2024okk}. The $\eta\Lambda$ elastic terms in
the external-mixing amplitude, Eq.~\eqref{eq:mixMB}, also
contribute to the line shape. Their interference with the other
production and rescattering terms is retained in the full
distribution.

{The comparison with the no-mixing results shows
noticeable changes in the $\pi^0\Lambda$ and
$\pi^0\bar\Lambda$ distributions. Without mixing, the
strength is more concentrated around the $\pi\Sigma$ threshold
region. The $\eta\Lambda$ and $\eta\bar\Lambda$
distributions also show stronger near-threshold enhancements.}

The $\Lambda\bar\Lambda$ invariant-mass distribution is broad and
follows the phase-space shape rather closely. The moderate
difference between the two curves results from the meson--meson
and meson--baryon interactions after integration over the
remaining phase-space variables.

\subsection{\texorpdfstring{$\psi(3686)\to\Lambda\bar\Lambda\pi^0\eta$}{The psi(3686) decay}}

\begin{figure}[htbp]
 \centering
   \includegraphics[width=0.96\textwidth]{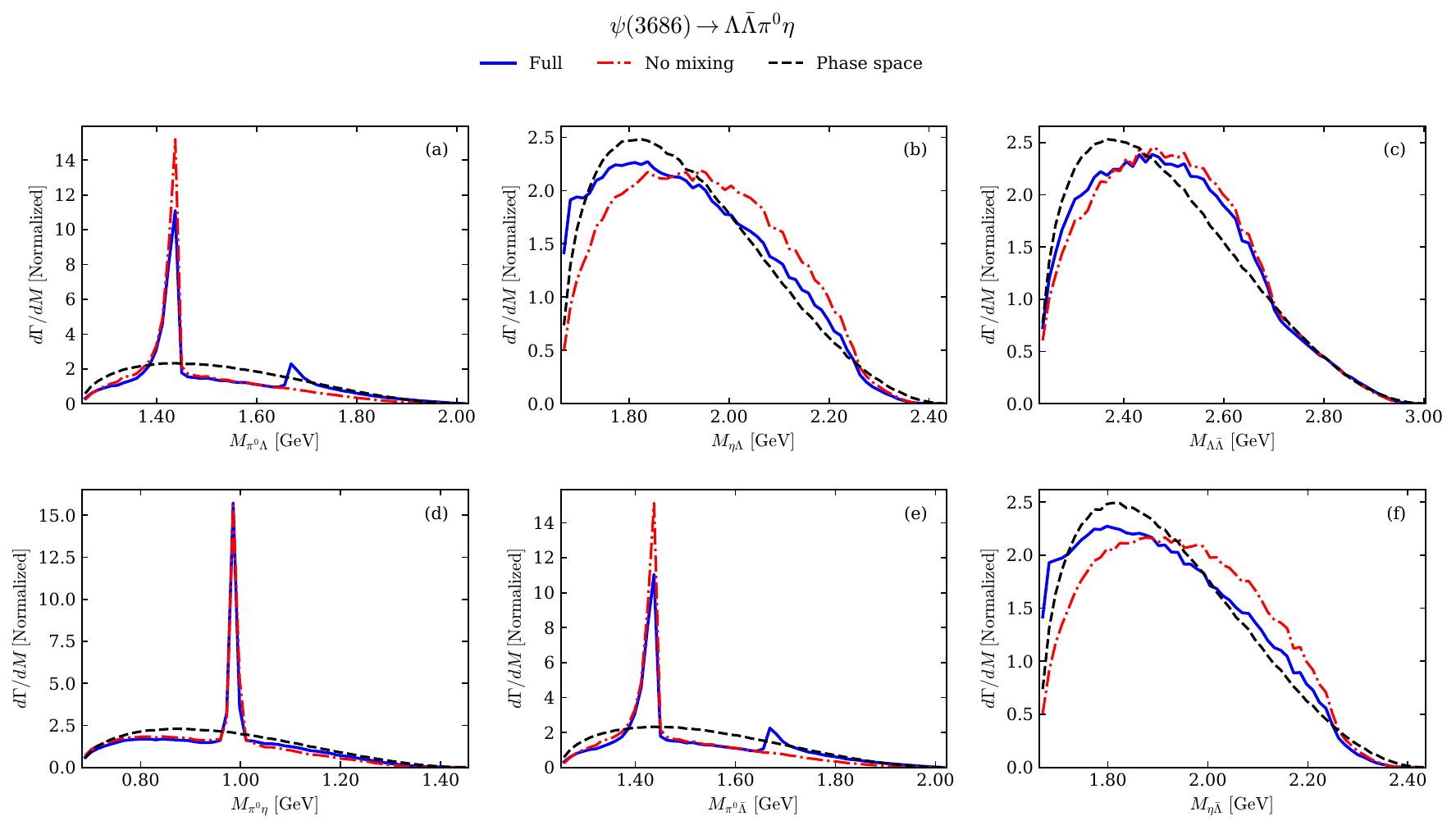}
 \caption{Same as Fig.~\ref{fig:jpsi}, but for
 $\psi(3686)\to\Lambda\bar\Lambda\pi^0\eta$. }
 \label{fig:psi}
\end{figure}

The larger mass of the $\psi(3686)$ provides a wider kinematic
region for the two-body subsystems. The corresponding upper limits
include
\begin{equation}
\begin{aligned}
 M_{\pi^0\Lambda}^{\max}&=2022.5\ \mathrm{MeV},\\
 M_{\pi^0\eta}^{\max}&=1454.7\ \mathrm{MeV}.
\end{aligned}
\label{eq:psi-limit}
\end{equation}
The $\pi^0\Lambda$ and $\pi^0\bar\Lambda$ distributions cover both
$\bar K N$ thresholds, and the $\pi^0\eta$ distribution extends
through the $a_0(980)$ region. The six calculated distributions
are shown in Fig.~\ref{fig:psi}.

The $M_{\pi^0\eta}$ distribution shows a pronounced peak
around $980$ MeV, corresponding to the $a_0(980)$ resonance,
which is generated by the coupled $\pi\eta$--$K\bar K$ interaction~\cite{Oller:1997ti,Lin:2021isc}.
{The peak remains prominent in the no-mixing
calculation, with a larger height in the normalized distribution.
This shows the role of the kaon mass difference in the
$\pi^0\eta$ spectrum.}

{{blue}The first thing is to note is the narrow width, unlike the normal width of the $a_0(980)$ is isospin allowed reactions~\cite{BESIII:2016tqo}.} 
This is the case because  the isospin forbidden reactions stem from the mass different between $K^+$ and $K^0$~\cite{Achasov:2002hg,Wu:2007jh,Hanhart:2007bd,Aceti:2012dj,Roca:2012cv,Li:2025rlj}

The $\pi^0\Lambda$ and $\pi^0\bar\Lambda$ distributions show
pronounced asymmetric enhancements around $1430$ MeV, followed
by a rapid decrease across the $\bar K N$ threshold region. These
structures are associated with the isovector coupled-channel
dynamics involving $\bar K N$, $\pi\Sigma$, and $\pi\Lambda$,
which gives rise to the $\Sigma(1430)$-region enhancement
\cite{Oller:2000fj,Jido:2003cb,Li:2024tvo,Li:2025yad}. 
This resonance has been observed by the Belle collaboration as a small peak over a background~\cite{Belle:2022ywa}. Here it stands as a peak with large strength compared with the rest of contributions.
{The enhancements near $1430$ MeV also appear in
the no-mixing distributions, with a larger peak height than in
the full calculation.}

A further structure appears around $1660$--$1670$ MeV in the
$\pi^0\Lambda$ and $\pi^0\bar\Lambda$ projections. This region
is close to the $\eta\Lambda$ threshold. The transition
$\eta\Lambda\to\pi^0\Lambda$ in the direct MB amplitude,
Eq.~\eqref{eq:tMB}, transfers the energy dependence of the
isoscalar $\eta\Lambda$ interaction to the $\pi^0\Lambda$
distribution. The external-mixing contribution in
Eq.~\eqref{eq:mixMB}, which contains the $\eta\Lambda$ elastic
amplitude, also contributes in this region (see diagrams to left in fig.~\ref{fig:mb-mix-eta}). These contributions
provide a connection between the $\Lambda(1670)$ dynamics and
the $\pi^0\Lambda$ invariant-mass spectrum.
{This structure becomes much weaker in the
no-mixing calculation. The comparison shows the important role
of $\pi^0$--$\eta$ mixing in producing the $\Lambda(1670)$
structure in the $\pi^0\Lambda$ and $\pi^0\bar\Lambda$
distributions.}

The $\eta\Lambda$ and $\eta\bar\Lambda$ distributions are broader,
with most of their strength in the region of approximately
$1700$--$1900$ MeV, followed by decreasing high-mass tails. The
$\Lambda\bar\Lambda$ distribution has a broad maximum around
$2400$ MeV and remains close to the phase-space shape. The
different structures in these projections reflect the combined
effects of the production weights, the coupled-channel
interactions, and the four-body kinematics.

\subsection{Comparison of the two decays}

The two charmonium decays provide complementary invariant-mass
regions for studying the final-state interactions. In the
$J/\psi$ decay, the $\pi^0\eta$ invariant mass remains below the
$a_0(980)$ region, and the $\pi^0\Lambda$ threshold is close to
the lower $\bar K N$ threshold. The most pronounced structure
therefore appears in the $\eta\Lambda$ and $\eta\bar\Lambda$
distributions near the $\Lambda(1670)$ region. In the
$\psi(3686)$ decay, the larger phase space makes the
$a_0(980)$ region in $\pi^0\eta$ and the $\bar K N$ threshold
region in $\pi^0\Lambda$ and $\pi^0\bar\Lambda$ accessible.
The $\Lambda(1670)$ dynamics also contributes to the
$\pi^0\Lambda$ and $\pi^0\bar\Lambda$ projections through the
$\eta\Lambda$ transition and external-mixing terms.

The six invariant-mass distributions are correlated through the
four-body kinematics. In particular, the energy dependence of a
two-body amplitude appears in more than one projection after
integration over the other phase-space variables.
The resulting spectra thus provide a combined picture of the meson--meson, meson--baryon, and meson--antibaryon interactions in the two decays.

\section{Conclusions}\label{sec:conclusions}

We have studied the four-body decays
$J/\psi,\psi(3686)\to\Lambda\bar\Lambda\pi^0\eta$ by combining
the SU(3) structure of the primary production with the
final-state interactions of meson--meson, meson--baryon, and
meson--antibaryon pairs. The two-body interactions are described
within the coupled-channel approach, and isospin violation is
introduced through the physical masses of the charged channels
and external $\pi^0$--$\eta$ mixing. These contributions to
the decay amplitude are summed coherently, and the six
two-particle invariant-mass distributions are obtained from the
four-body phase space.

The calculated distributions exhibit structures associated with
three important coupled-channel regions. The $\pi^0\eta$
spectrum in the $\psi(3686)$ decay shows a narrow enhancement
around the $K\bar K$ thresholds, corresponding to the
$a_0(980)$ dynamics. The $\pi^0\Lambda$ and
$\pi^0\bar\Lambda$ spectra show asymmetric enhancements around
the $\bar K N$ thresholds in the $\Sigma(1430)$ region. The
$\eta\Lambda$ and $\eta\bar\Lambda$ distributions develop
enhancements close to threshold, reflecting the isoscalar
interaction associated with the $\Lambda(1670)$. The latter
dynamics is also visible in the $\pi^0\Lambda$ and
$\pi^0\bar\Lambda$ spectra through the transition and
external-mixing amplitudes.

The different masses of $J/\psi$ and $\psi(3686)$ allow us
to explore these interactions in different kinematic regions.
 The $J/\psi$ decay emphasizes the
near-threshold $\eta\Lambda$ and $\eta\bar\Lambda$ structures,
whereas the larger phase space in the $\psi(3686)$ decay also
allows the $a_0(980)$ and $\Sigma(1430)$ regions to be explored.
{The no-mixing results show that the
mass differences among the charge channels also contribute to
the invariant-mass distributions. The effect of $\pi^0$--$\eta$
mixing differs between the two decays because of the available
phase space.}
The six invariant-mass distributions illustrate how the
coupled-channel dynamics of different two-body subsystems,
together with their interference and the four-body kinematics,
shape the observable spectra. These results provide a basis for
future studies of isospin violation and hadron interactions in
charmonium four-body decays.
We should also stress that the signals that we have observed due to final state interaction correspond to the production of dynamically generated resonances which stem from the interaction of coupled channels. Genuine resonances that we would attribute to three quark states, like the $\Sigma(1385)$ should not appear in these isospin violating reactions, a feature that the experiment should prove.
As stated in~\cite{Ren:2026ila}, these isospin violating reactions are a useful tool to separate genuine resonances from those that are of molecular type.

\section*{Acknowledgments}
We would like to thank Dr. Wen-Tao Lyu for very useful discussions.
This work is partly supported by the
National Natural Science Foundation of China (NSFC)
under Grants No. 12405089, and by the China Postdoctoral Science Foundation under Grant No. 2022M720360
and No. 2022M720359. J. Song would also like to thank the support from the Hainan Provincial Excellent Talent Team under the “Four Talents” Gathering Program of Hainan Province.
Yu-Shan Ren acknowledges the support from the China Scholarship Council.
This work is also supported by the Grants PID2023-
147458NB-C21 and CEX2023-001292-S funded by MI-
CIU/AEI/10.13039/501100011033 and by ERDF/EU,
as well as of the PROMETEO program Grant
CIPROM/2023/59 funded by Generalitat Valenciana
10.13039/501100003359.

\section*{Appendix: Evaluation of phase space}

\subsubsection{Reduction of the four-body phase space}
We provide details in the evaluation of $ \Gamma_\Psi$ of Eq.(~\ref{eq:width}). Recall we have the indices $\Lambda(p_1)\,\pi^0(p_2)\,\eta(p_3)\,\bar\Lambda(p_4)$. We follow closely the formalism of Ref.~\cite{Song:2022kac}
We work in the rest frame of the initial charmonium,
$Q=(M_\Psi,\bm 0)$. The three-momentum delta function in
Eq.~\eqref{eq:phase-space} fixes
\begin{equation}
 \bm p_4=-\bm p_1-\bm p_2-\bm p_3.
 \label{eq:p4-momentum}
\end{equation}
For the two mesons, it is convenient to introduce their total and
relative three-momenta,
\begin{equation}
 \bm P_m=\bm p_2+\bm p_3,\qquad
 \bm q=\bm p_2-\bm p_3.
 \label{eq:Pq-def}
\end{equation}
The individual momenta are then
\begin{equation}
 \bm p_2=\frac{\bm P_m+\bm q}{2},\qquad
 \bm p_3=\frac{\bm P_m-\bm q}{2},\qquad
 \bm p_4=-\bm p_1-\bm P_m,
 \label{eq:Pq}
\end{equation}
with the Jacobian
\begin{equation}
 \dd^3\bm p_2\,\dd^3\bm p_3
 =\frac{1}{8}\dd^3\bm P_m\,\dd^3\bm q.
 \label{eq:jacobian-Pq}
\end{equation}
After eliminating $\bm p_4$, the phase-space element becomes
\begin{align}
 \dd\Phi_4={}&\frac{1}{(2\pi)^8}
 \frac{\dd^3\bm p_1}{2E_1}
 \frac{\dd^3\bm p_2}{2\omega_\pi}
 \frac{\dd^3\bm p_3}{2\omega_\eta}
 \frac{1}{2E_4}
 \nonumber\\
 &\times
 \delta(M_\Psi-E_1-\omega_\pi-\omega_\eta-E_4),
 \label{eq:phase-space-after-p4}
\end{align}
where $E_1=\sqrt{m_\Lambda^2+\bm p_1^2}$ and
$E_4=\sqrt{m_\Lambda^2+\bm p_4^2}$.

We choose the $z$ axis along $\bm p_1$ and write
\begin{equation}
 \bm p_1=p\,\hat{\bm z},\qquad
 p=\sqrt{E_1^2-m_\Lambda^2}.
 \label{eq:p1-axis}
\end{equation}
The direction of $\bm P_m$ is specified by the angles
$(\theta,\phi)$ with respect to the $z$ axis.
The direction of $\bm q$ is specified by
$(\widetilde\theta_q,\widetilde\phi_q)$ in a coordinate system
with its $z$ axis along $\bm P_m$.
Thus,
\begin{equation}
 \bm P_m=P_m
 \begin{pmatrix}
  \sin\theta\cos\phi\\
  \sin\theta\sin\phi\\
  \cos\theta
 \end{pmatrix},\qquad
 \widetilde{\bm q}=q
 \begin{pmatrix}
  \sin\widetilde\theta_q\cos\widetilde\phi_q\\
  \sin\widetilde\theta_q\sin\widetilde\phi_q\\
  \cos\widetilde\theta_q
 \end{pmatrix}.
 \label{eq:Pm-q-directions}
\end{equation}
The momentum $\bm q$ in the original coordinate system is obtained
through the rotation
\begin{equation}
 \bm q=R_z(\phi)R_y(\theta)\widetilde{\bm q},
 \label{eq:q-rotation}
\end{equation}
where
\begin{equation}
 R_z(\phi)R_y(\theta)=
 \begin{pmatrix}
  \cos\phi\cos\theta&-\sin\phi&\cos\phi\sin\theta\\
  \sin\phi\cos\theta& \cos\phi&\sin\phi\sin\theta\\
  -\sin\theta&0&\cos\theta
 \end{pmatrix}.
 \label{eq:rotation-matrix}
\end{equation}
This choice of variables fixes the relative direction of the two
meson momenta and allows us to construct the four final-state
momenta.

The meson energies follow directly from Eq.~\eqref{eq:Pq},
\begin{align}
 \omega_\pi&=
 \sqrt{m_\pi^2+\frac{P_m^2+q^2}{4}
       +\frac{P_mq}{2}\cos\widetilde\theta_q},
 \nonumber\\
 \omega_\eta&=
 \sqrt{m_\eta^2+\frac{P_m^2+q^2}{4}
       -\frac{P_mq}{2}\cos\widetilde\theta_q}.
 \label{eq:energies}
\end{align}
The on-shell condition for the final-state antibaryon gives
\begin{equation}
 E_4^2=m_\Lambda^2+p^2+P_m^2
       +2pP_m\cos\theta.
 \label{eq:E4-angle}
\end{equation}
Combining this relation with energy conservation,
$E_4=M_\Psi-E_1-\omega_\pi-\omega_\eta$, we obtain
\begin{equation}
 \cos\theta=A=
 \frac{(M_\Psi-E_1-\omega_\pi-\omega_\eta)^2
       -m_\Lambda^2-p^2-P_m^2}{2pP_m}.
 \label{eq:angle}
\end{equation}
We integrate over $\cos\theta$ using the energy-conservation
delta function. With
$\left|\partial E_4/\partial\cos\theta\right|=pP_m/E_4$,
we obtain
\begin{align}
 &\int_{-1}^{1}\frac{\dd\cos\theta}{2E_4}
 \delta(M_\Psi-E_1-\omega_\pi-\omega_\eta-E_4)
 \nonumber\\
 &\qquad =
 \frac{1}{2pP_m}
 \Theta(1-A^2)
 \Theta(M_\Psi-E_1-\omega_\pi-\omega_\eta).
 \label{eq:angular-delta}
\end{align}

After integrating over $\cos\theta$ and using the Jacobian
in Eq.~\eqref{eq:jacobian-Pq}, we obtain the decay width
as
\begin{align}
 \Gamma_\Psi={}&
 \frac{(2m_\Lambda)^2}{2M_\Psi}
 \frac{1}{16(2\pi)^7}
 \int \dd E_1\,P_m\dd P_m\,\dd\phi
 \nonumber\\
 &\times q^2\dd q\,
 \dd\cos\widetilde\theta_q\,\dd\widetilde\phi_q\,
 \frac{2|t_{\mathrm{total}}|^2}
      {(2\omega_\pi)(2\omega_\eta)}
 \nonumber\\
 &\times\Theta(1-A^2)
 \Theta(M_\Psi-E_1-\omega_\pi-\omega_\eta).
 \label{eq:reduced-width}
\end{align}
The integration variables are
$E_1$, $P_m$, $q$, $\phi$, $\cos\widetilde\theta_q$, and
$\widetilde\phi_q$.

\subsubsection{Integration limits and invariant-mass distributions}

For the numerical evaluation of Eq.~\eqref{eq:reduced-width},
the integration variables are in the ranges
\begin{equation}
\begin{aligned}
 m_\Lambda&\leq E_1\leq E_{1,\max},&
 0&\leq P_m\leq P_{m,\max},\\
 0&\leq q\leq q_{\mathrm{box}},&
 -1&\leq\cos\widetilde\theta_q\leq1,\\
 0&\leq\phi<2\pi,&
 0&\leq\widetilde\phi_q<2\pi,
\end{aligned}
\label{eq:integration-ranges}
\end{equation}
where
\begin{align}
 E_{1,\max}
 &=\frac{M_\Psi^2+m_\Lambda^2
 -(m_\Lambda+m_\pi+m_\eta)^2}{2M_\Psi},
 \nonumber\\
 P_{m,\max}
 &=\frac{\sqrt{\lambda\!\left(
 M_\Psi^2,(m_\pi+m_\eta)^2,4m_\Lambda^2\right)}}{2M_\Psi}.
 \label{eq:bounds}
\end{align}
Here $\lambda(x,y,z)=x^2+y^2+z^2-2xy-2xz-2yz$ is the
Källén function. 
The upper limit of $E_1$ corresponds to the minimum invariant
mass of the $\pi^0\eta\bar\Lambda$ system, while
$P_{m,\max}$ is determined by the minimum invariant masses
of the $\pi^0\eta$ and $\Lambda\bar\Lambda$ pairs.

The upper limit of the relative momentum $q$ is determined from
the physical four-body kinematics. At fixed
$M_{23}=M_{\pi^0\eta}$ and $M_{14}=M_{\Lambda\bar\Lambda}$,
the energy and momentum of the meson pair in the charmonium
rest frame are
\begin{equation}
 E_{23}=\frac{M_\Psi^2+M_{23}^2-M_{14}^2}{2M_\Psi},
 \qquad
 P_{23}=\frac{\sqrt{\lambda(M_\Psi^2,M_{23}^2,M_{14}^2)}}{2M_\Psi}.
 \label{eq:pair-kinematics}
\end{equation}
In the $\pi^0\eta$ rest frame, the relative momentum and the
energy difference of the two mesons are
\begin{equation}
 k^*=\frac{\sqrt{\lambda(M_{23}^2,m_\pi^2,m_\eta^2)}}{2M_{23}},
 \qquad
 E_\pi^*-E_\eta^*=\frac{m_\pi^2-m_\eta^2}{M_{23}}.
 \label{eq:meson-pair-rest}
\end{equation}
Maximizing $|\bm p_2-\bm p_3|$ over the direction
of the meson momenta, we obtain
\begin{equation}
 q_{\max}(M_{23},M_{14})=
 \frac{2E_{23}k^*+P_{23}|E_\pi^*-E_\eta^*|}{M_{23}},
 \label{eq:qmax-boost}
\end{equation}
or, equivalently,
\begin{equation}
 q_{\max}(M_{23},M_{14})=
 \frac{
 (M_\Psi^2+M_{23}^2-M_{14}^2)
 \sqrt{\lambda(M_{23}^2,m_\pi^2,m_\eta^2)}
 +\sqrt{\lambda(M_\Psi^2,M_{23}^2,M_{14}^2)}
 |m_\pi^2-m_\eta^2|}
 {2M_\Psi M_{23}^2}.
 \label{eq:qmax}
\end{equation}
The second term arises from the mass difference between
the pion and eta mesons.
Here the maximization is performed over the physical region
$m_\pi+m_\eta\leq M_{23}\leq M_\Psi-2m_\Lambda$,
$2m_\Lambda\leq M_{14}\leq M_\Psi-m_\pi-m_\eta$, and
$M_{23}+M_{14}\leq M_\Psi$. 
We obtain $q_{\max}\simeq 507.4$ MeV for $J/\psi$ and
$q_{\max}\simeq 1244.3$ MeV for $\psi(3686)$.
Accordingly, we choose
\begin{equation}
 q_{\mathrm{box}}=
 \begin{cases}
 520~\mathrm{MeV}, & \Psi=J/\psi,\\
 1250~\mathrm{MeV}, & \Psi=\psi(3686).
 \end{cases}
 \label{eq:qbox}
\end{equation}
These values cover the allowed kinematic regions.
We perform the numerical integration by randomly sampling
the six variables within the specified ranges.
The physical events are selected by the step functions in
Eq.~\eqref{eq:reduced-width}, and their contributions are
weighted by the corresponding integrand.

For each accepted Monte Carlo event, we determine the angle
$\theta=\arccos A$ and calculate the final-state momenta using
Eqs.~\eqref{eq:Pm-q-directions}--\eqref{eq:q-rotation}
and Eq.~\eqref{eq:Pq}.
 The six invariant masses are then calculated
from
\begin{equation}
 M_{ij}^2=(E_i+E_j)^2-
          |\bm p_i+\bm p_j|^2.
 \label{eq:invariant-mass-reconstruction}
\end{equation}
Their kinematic limits are
\begin{equation}
 m_i+m_j\leq M_{ij}\leq M_\Psi-m_k-m_l,
 \qquad \{i,j,k,l\}=\{1,2,3,4\}.
 \label{eq:mass-bounds}
\end{equation}

For each invariant mass, we sum the event weights in each bin and divide by the bin width to calculate the mass distribution.

\newpage

\bibliography{refs} 
\end{document}